\documentclass[aps,prapplied,reprint,amsmath,amssymb,superscriptaddress,floatfix]{revtex4-2}
\usepackage[dvips]{graphicx}
\usepackage{%
qcircuit,
physics,
hyperref,
siunitx,
xcolor}

\hypersetup{
    colorlinks,
    linkcolor={blue!50!black},
    citecolor={blue!50!black},
    urlcolor={blue!80!black}
}

\usepackage[normalem]{ulem}
\usepackage[capitalize]{cleveref}

\begin{document}


\title{
Multicore Quantum Computing
}

\newcommand{\qmaddress}{Quantum Motion, 9 Sterling Way, London N7 9HJ, United Kingdom}

\author{Hamza Jnane}
\email{hamza.jnane@materials.ox.ac.uk}
\affiliation{\qmaddress}
\affiliation{Department of Materials, University of Oxford, Parks Road, Oxford OX1 3PH, United Kingdom}

\author{Brennan Undseth}
\affiliation{\qmaddress}

\author{Zhenyu Cai}
\affiliation{\qmaddress}
\affiliation{Department of Materials, University of Oxford, Parks Road, Oxford OX1 3PH, United Kingdom}

\author{Simon C Benjamin}
\affiliation{\qmaddress}
\affiliation{Department of Materials, University of Oxford, Parks Road, Oxford OX1 3PH, United Kingdom}

\author{B\'alint Koczor}
\email{balint.koczor@materials.ox.ac.uk}
\affiliation{\qmaddress}
\affiliation{Department of Materials, University of Oxford, Parks Road, Oxford OX1 3PH, United Kingdom}


\begin{abstract}
Any architecture for practical quantum computing must be scalable. An attractive approach is to create multiple {\it cores}, computing regions of fixed size that are well-spaced but interlinked with communication channels. This exploded architecture can relax the demands associated with a single monolithic device: the complexity of control, cooling and power infrastructure as well as the difficulties of cross-talk suppression and near-perfect component yield. Here we explore interlinked multicore architectures through analytic and numerical modelling. While elements of our analysis are relevant to diverse platforms, our focus is on semiconductor electron spin systems in which numerous cores may exist on a single chip within a single fridge. We model shuttling and microwave-based interlinks and estimate the achievable fidelities, finding values that are encouraging but markedly inferior to intra-core operations. We therefore introduce optimised entanglement purification to enable high-fidelity communication, finding that $99.5\%$ is a very realistic goal. We then assess the prospects for quantum advantage using such devices in the NISQ-era and beyond: we simulate recently proposed exponentially-powerful error mitigation schemes in the multicore environment and conclude that these techniques impressively suppress imperfections in both the inter- and intra-core operations.
\end{abstract}

\maketitle

\section{Introduction and Overview}
It is widely believed that the era of practical quantum computing is now dawning. 
With an increasing tempo, new records for qubit count are set and then surpassed~\cite{aruteQuantumSupremacyUsing2019,zhongPhaseProgrammableGaussianBoson2021,tangCutQCUsingSmall2021,wuStrongQuantumComputational2021,ebadiQuantumPhasesMatter2021,IBMUnveilsBreakthrough2021,gongQuantumWalksProgrammable2021a}.
However, as any given platform scales it is likely that there will be critical device sizes that will prove difficult to scale beyond. For example,
it is generally expected that linear ion traps will be difficult to scale beyond the 50-100 ion scale due to decrystallisation, spectral crowding etc.~\cite{paganoCryogenicTrappedionSystem2018a,cetinaControlTransverseMotion2022,landsmanTwoqubitEntanglingGates2019}.
Moreover, as monolithic (single-chip and single-core) solid-state arrays scale they face increasing challenges in providing power, control or cooling infrastructure to the qubit lattice. Furthermore, cross-talk and frequency crowding become increasingly problematic and there is a growing probability that a crucial component somewhere within the array will have a fabrication error (due to finite yield). It is to be expected that it will prove possible to create a plurality of distinct, moderate-scale devices -- a multicore system -- long before monolithic structures become truly scalable.

\begin{figure*}[tb]
	\begin{centering}
		\includegraphics[width=0.7\textwidth]{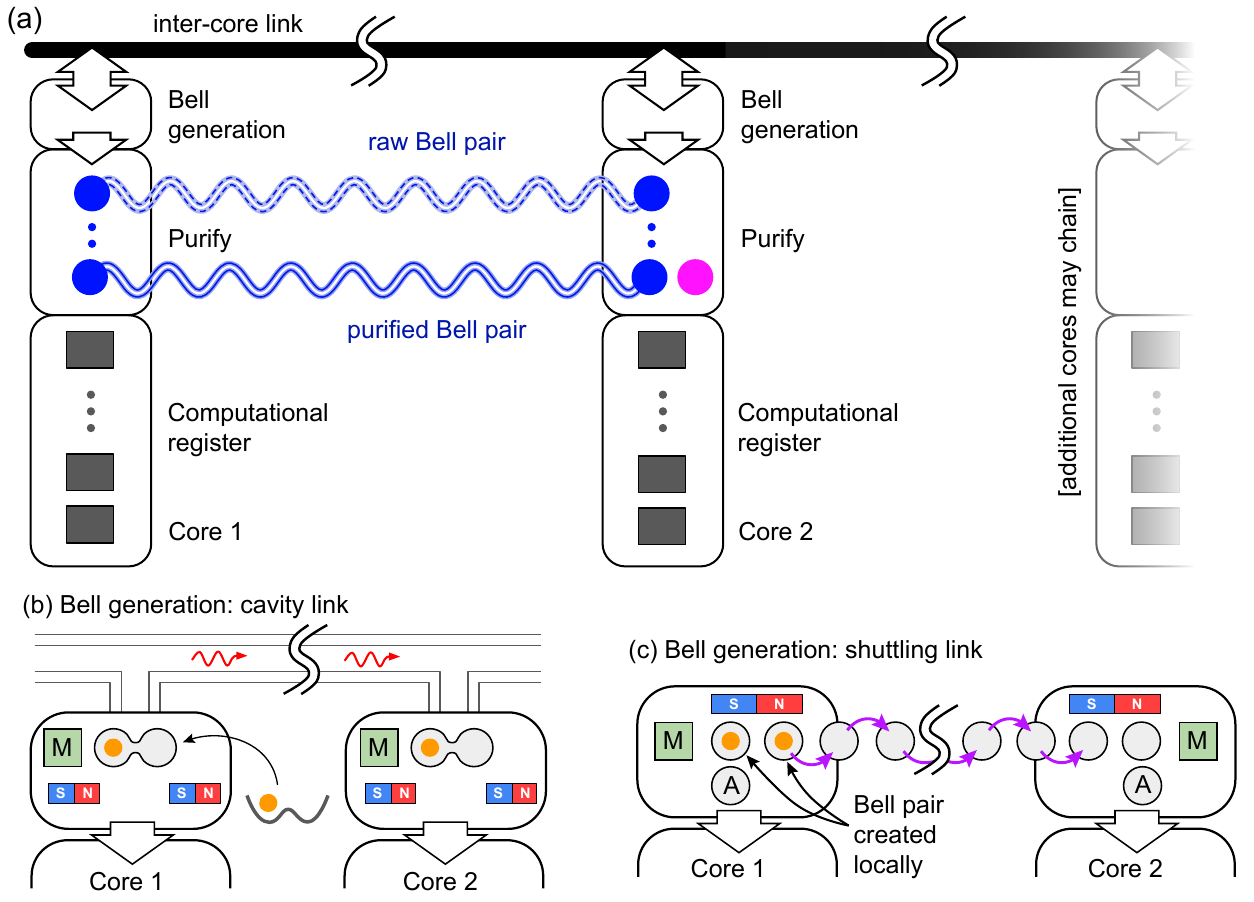}
		\caption{(a) Macroscopically separate quantum cores (core 1, core 2 etc.) are linked via purified Bell pairs (wavy lines). Such an architecture is well-suited for implementing the ESD/VD error mitigation technique \cite{koczorExponentialErrorSuppression2021, hugginsVirtualDistillationQuantum2021}: multiple quantum cores perform the same quantum computation on $N$ qubits and these copies are used to verify each other via a derangement operation. This exponentially reduces the impact of imperfections in both the local operations and in the Bell pairs. The derangement operation is implemented by teleporting individual qubits into a buffer qubit (magenta dot) in one of the cores thus consuming overall $N$ bell pairs.
		In this work we consider distributing Bell pairs via a cavity (b) and by shuttling (c) (not to scale).
		(b)	Schematic of two spin registers connected via a superconducting cavity. By hybridising the spin and charge degrees of freedom of electrons in DQDs, we can couple distant qubits through microwave photons (red squiggles) in a cavity. Charge measurements are used to speed up the process. A successful measurement outcome will lead to a raw Bell pair, whose quality can be improved through distillation. 
		(c) Schematic of two spin registers connected via a quantum dot shuttling channel. A Bell pair consisting of two electrons (small orange dots) is created locally via high-fidelity local gates; One of the electrons is then shuttled through the few micron-length chain on a \SI{100}{\nano\second} timescale
		--
    	this chain does not use micromagnets given we need not control the spin degree of freedom during shuttling.
			\label{fig:ill}
		}
	\end{centering}
\end{figure*}

The multicore approach may be a key evolutionary stage  even if the long-term goal is to realise a massive monolithic array of qubits. The progression is natural: a single core is realised (arguably, the `supremacy' class devices already reported are of this kind~\cite{aruteQuantumSupremacyUsing2019,wuStrongQuantumComputational2021,IBMUnveilsBreakthrough2021}); multiple cores are realised without quantum interlinks (achieving simple parallelism that is already valuable for sampling-based tasks including variational quantum algorithms~\cite{caiResourceEstimationQuantum2020}); then multiple cores are realised with interlinks as discussed in this paper. Subsequently as cores become larger and interlinks more dense, the progression reaches large-scale regular lattice patterns that constitute the fabric for topological error correcting codes~\cite{fowlerSurfaceCodesPractical2012,li2DCompassCodes2019,chamberlandTopologicalSubsystemCodes2020}.

We begin by considering the hardware-level realisation of the interlinks required for the multicore concept. Specifically, we consider single-electron qubits in semiconductor (e.g. silicon) devices and study two promising solutions: electron shuttling and microwave-based remote interactions, the latter operated in the fast (resonant) regime. While other solutions have been suggested, notably optical links~\cite{heTwoqubitGatePhosphorus2019,higginbottomOpticalObservationSingle2022}, we find that these two linking modes have the potential to realise strong entanglement over relevant length scales when we assume that components perform at current or near-future levels. In particular, whereas shuttling is well suited for linking multiple cores on a single chip, cavities are potentially useful both for intra-chip and inter-chip linkage. It is worth noting that other approaches to the multicore concept on different platforms have been studied ~\cite{meteriiiArchitectureQuantumMulticomputer2006, jiangDistributedQuantumComputation2007, oiScalableErrorCorrection2006}.

Although we consider our models' predictions of interlink performance to be encouraging, the fidelities that we obtain are well below those expected of short-range, intra-core operations. One might regard this as a fundamental constraint given the lessons of classical computing: a very mature technology in which links are inevitably inferior over each scale of distance. Intra-CPU operations can be an order of magnitude faster than calls to memory subsystems, which in turn are an order of magnitude faster than fibre-based links between adjacent `boxes' in a cluster. It is therefore reasonable to assume that longer range links may always be inferior to local links, even as quantum computing matures. Our modelling suggests this is true of the fidelity, rather than the speed, of inter- versus intra-core links. This was anticipated as early as 2003 by \textcite{durEntanglementPurificationQuantum2003} who proposed the use of efficient entanglement purification to upgrade the channel fidelity: Low quality Bell pairs are distributed via the channel, and then combined using local operations to yield high-fidelity pairs; each such pair can be consumed to enact a desired two-qubit remote gate. Schemes of this kind have been much discussed ~\cite{durEntanglementPurificationQuantum2003,krastanovOptimizedEntanglementPurification2019,nickersonFreelyScalableQuantum2014,lopiparoResourceReductionDistributed2020,nigmatullinMinimallyComplexIon2016} and indeed have been experimentally realised~\cite{kalbEntanglementDistillationSolidstate2017}. We adopt an approach of this kind, purifying Bell pairs to create the resource needed for high-fidelity remote operations using an architecture shown schematically in \cref{fig:ill}. By suitably tailoring purification protocols we find fidelities (with respect to the closest of the four canonical Bell states) of $99.5\%$ can be achieved after a few successive uses of the interlink. Higher values are achievable in principle, but as anticipated in~\cite{durEntanglementPurificationQuantum2003} the noise in intra-core operations becomes the limitation. 

Having thus estimated the achievable speed-fidelity curve for the interlinks, we conclude by examining the potential of this multicore architecture for achieving quantum advantage. While we recognise the value for long-term, fault tolerant quantum computing by distributing an error correcting code over the cores, we focus on a NISQ-era application: A recently discovered error mitigation technique that allows one to suppress errors as an exponentially decreasing function of hardware size. The idea was introduced in Ref.~\cite{koczorExponentialErrorSuppression2021} as `Error Suppression by Derangements' (ESD) and in Ref.~\cite{hugginsVirtualDistillationQuantum2021} as `Virtual Distillation' (VD). In the present paper we model the performance of two interlinked cores and present it in the context of the in-principle performance of three- and four-core systems. Our conclusion is that the approach should be profoundly enabling for NISQ-era tasks and the prospects of real quantum advantage in the sense of meaningful tasks that are infeasible on classical systems.

\section{Distributing Bell pairs using cavities}\label{sec:cavity_mediated_gate}
In this section we first consider the possibility of generating Bell pairs using cavity mediated interactions. We explicitly model a specific setup assuming parameters that are achievable with near-term hardware and conclude that fidelities on the order of $95\%$ are realistic. Our numerical models also confirm that optimised purification protocols can increase this fidelity to levels that are comparable to intra-core operations.

\subsection{Overview}

In the context of superconducting qubits, the coupling of distant qubits can be achieved by mediating quantum information through a superconducting cavity within the circuit quantum electrodynamics (cQED) framework \cite{blaisCavityQuantumElectrodynamics2004,blaisQuantuminformationProcessingCircuit2007,majerCouplingSuperconductingQubits2007,sillanpaaCoherentQuantumState2007}. By using this method, one can generate a coherent coupling between the resonator's modes and the qubits over few millimetres. For simplicity and for consistency with the shuttling alternative considered presently, we will restrict ourselves to single-spin qubits while recognising that singlet-triplet qubits may have advantages in cavity coupling \cite{jinStrongCouplingSpin2012}.

Despite their long coherence time and the possibility of manufacturing them with current industry standards \cite{maurandCMOSSiliconSpin2016,veldhorstSiliconCMOSArchitecture2017, gonzalez-zalbaScalingSiliconbasedQuantum2021, burkardSemiconductorSpinQubits2021}, applying the above procedure to single-electron silicon spin qubits is hard. The reason is that electronic spins have a small magnetic dipole, which means that the spin-photon magnetic dipole coupling is slower than the spin's decoherence rate. Ref.~\cite{zhengCircuitQuantumElectrodynamics2021} gives a comprehensive review of cQED with single electron spin qubits. 

In recent years numerous experimental works have explored the possibility of coherently coupling a semiconductor spin qubit with microwave photons~\cite{miCoherentSpinPhoton2018,samkharadzeStrongSpinphotonCoupling2018,landigCoherentSpinPhoton2018}. By placing a cobalt micromagnet above a single electron confined in a double quantum dot (DQD), references~\cite{miCoherentSpinPhoton2018,samkharadzeStrongSpinphotonCoupling2018} have leveraged the following two-step process to effectively couple electron spins with a cavity. First, the large magnetic field gradient generated by the micromagnet increases the spin-orbit coupling which leads to a hybridisation of the spin and charge degrees of freedom. Then, the dipole-dipole interaction between the cavity and the charge indirectly couples the spin to the cavity resulting in significant coupling rate (on the order of several MHz). 

Despite the fact that hybrid spin qubits are more prone to charge noise, references~\cite{miCoherentSpinPhoton2018,samkharadzeStrongSpinphotonCoupling2018} have achieved the strong coupling regime, i.e., the spin-photon coupling rate is greater than the spin decoherence rate and the cavity loss rate. 
Recently, Borjans et al. \cite{borjansResonantMicrowavemediatedInteractions2020}, demonstrated that one can coherently couple two spin qubits with a resonator by extending the architecture in \cite{miCoherentSpinPhoton2018} and Harvey-Collard et al. \cite{harvey-collardCoherentSpinSpinCoupling2022} reported the coupling of distant spins through virtual photons. The latter represents a significant step towards the experimental realisation of cavity-mediated two-qubit gates.

The success of these experiments has stimulated further theoretical work towards implementing long-range two-qubit gates. While  Ref.~\cite{benitoOptimizedCavitymediatedDispersive2019} showed how an $\text{iSWAP}$ gate can be implemented with fidelities above $90\%$ by optimising the spin-charge hybridisation with current hardware, Warren et al.~\cite{warrenLongdistanceEntanglingGates2019} have reported $99.5\%$ fidelity $\sqrt{\text{iSWAP}}$ gates by exploiting the low-energy dynamics of the system. More recently, the authors proposed a photon-mediated cross-resonance gate that is more robust to charge noise \cite{warrenRobustPhotonmediatedEntangling2021}. It is important to note that all these works assume a dispersive parameter regime: this regime may reduce cavity loss, but it results in relatively slow gates.

 In the present work we limit ourselves to generating specific Bell states using cavity mediated interactions,
in contrast to long-range general unitary quantum gates in the dispersive regime which were the focus of the aforementioned prior works ~\cite{warrenLongdistanceEntanglingGates2019,benitoOptimizedCavitymediatedDispersive2019}.
 This allows us to work in a resonant parameter regime that generates entanglement on a significantly faster timescale at the cost of introducing small (coherent) errors. We propose to apply powerful purification techniques to reduce such errors in our raw Bell pairs thereby non-deterministically generating high-fidelity Bell pairs. After purification, these Bell pairs can be used as a resource to perform arbitrary two-qubit gates via gate teleportation. As we explain below in \cref{sec:applications}, more generally these Bell pairs can be used as valuable resources for a range of specific practical tasks beyond gate teleportation.

\begin{figure}[tb]
	\begin{centering}
		\includegraphics[width=0.4\textwidth]{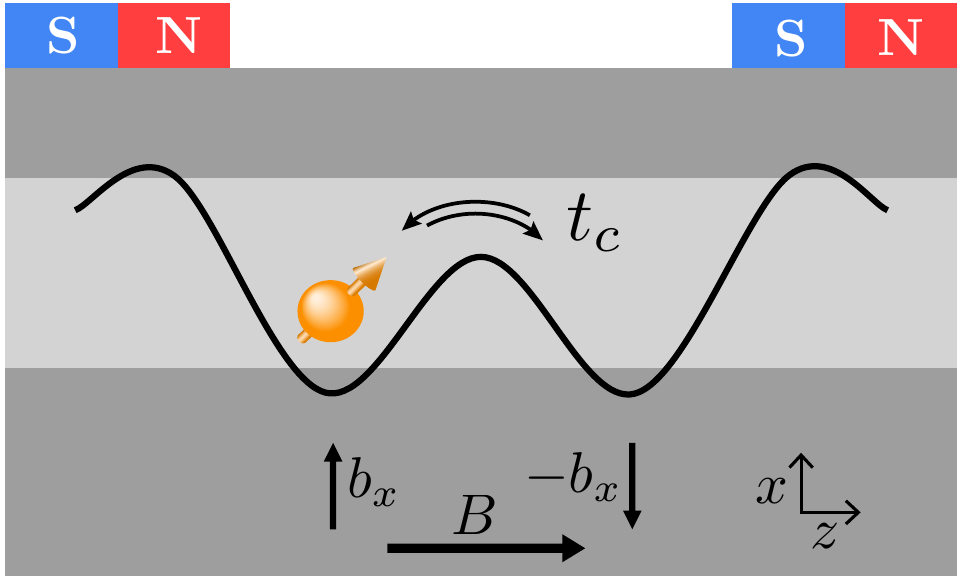}
		\caption{Sketch of a single electron confined in a DQD. The DQD is generated by gates on top of a material heterostructure (a Si/SiGe heterostructure for instance). Cobalt micromagnets are placed above the DQD to generate the transverse magnetic field $b_x$ which will hybridise the charge and spin degrees of freedom. By lifting the spin degeneracy, the parallel magnetic field $(0,0,B)^T$ allows us to define a qubit on the electron's spin. 
		}
		\label{fig:DQD}
	\end{centering}
\end{figure}

\subsection{The model}

By confining a single electron in a DQD, we obtain a \emph{spin qubit} encoded in the spin of the electron with the basis states $\ket{\uparrow}, \ket{\downarrow}$ and a \emph{charge qubit} encoded in the position (orbit) of the electron with the basis states $\ket{\text{L}}, \ket{\text{R}}$.
The spin and charge qubits are coupled by a magnetic field gradient across the two dots.
Let us note that in our modelling we do not consider higher excited states of the charge degree of freedom nor the possible coupling with the valley degree of freedom \cite{culcerInterfaceRoughnessValleyorbit2010}.
To first order these would not change our results in the present section due to other mechanisms of more significant imperfections as explained below (while we will need to take into account the valley coupling for the shuttling-based alternative).
As such, the full Hamiltonian of this system can be written as
\begin{equation}
    \begin{aligned}\label{eqn:DQD_ham}
    H_{\mathrm{DQD}} &=  \left(\frac{\epsilon}{2}\tau_{z} + t_c\tau_{x}\right) + \frac{1}{2} B s_{z} + \frac{1}{2} \tau_{z}\otimes (\vec{b} \cdot \vec{s}\,).
\end{aligned}
\end{equation}
Above and throughout the paper we set $\hbar = 1$ and express energy in terms of angular frequency units unless stated otherwise. Here $\tau_i$ and $s_i$ are Pauli matrices $\sigma_i$ with $i \in \{ x,y,z\}$ acting on the charge and spin qubits, respectively. The first term above represents the Hamiltonian for a charge qubit with $\epsilon$ and $t_c$ being the detuning and tunnel coupling between the two dots. The second term represents the Zeeman splitting $B$ of the spin qubit due to a magnetic field that defines the spin $z$ direction. The last term represents the coupling between the spin and charge qubits due to the \emph{difference} $\vec{b}:=(b_x, b_y, b_z)^T$ in magnetic fields between the two dots. In this section, we assume the main gradient of the magnetic field is along the $x$-direction ($\vec{b} = (b_x,0,0)^T$). This gradient can be generated using micromagnets as illustrated in \cref{fig:DQD}. While using micromagnets in coupling spins to cavities may present technical challenges \cite{borjansResonantMicrowavemediatedInteractions2020, harvey-collardCoherentSpinSpinCoupling2022}, we expect it is not a fundamental bottleneck for the cavity-based approach given only two micromagnets per pair of connected cores need to be implemented.

In our modelling we consider a resonator described by the usual quantum-harmonic-oscillator Hamiltonian $H_r = \omega_r a^{\dagger} a$ where $\omega_r$ is the resonator frequency and $\hat{a}^{\dagger}$, $\hat{a}$ are the corresponding (bosonic) creation and annihilation operators. Furthermore, we model the coupling between the resonator and the charge qubit via the Hamiltonian
\begin{equation}
    H_{\tau r} = g_c\tau_{z}\otimes \left(a+a^{\dagger}\right),
\end{equation}
where $g_c$ is the charge-photon dipole coupling factor. Hence, via interactions mediated through the charge qubit, we can couple two spin qubits in two distant DQDs via the resonator. We write the full Hamiltonian for such a system as
\begin{equation}
    H_{srs} = H_{r} +\sum_{i=1}^2 (H_{\mathrm{DQD},i} + H_{\tau r,i}),
    \label{eq:two_qubit_hamiltonian}
\end{equation}
where $i \in \{ 1,2 \}$ are labels of the two distant DQDs, each containing a spin qubit and a charge qubit. The resonant condition corresponds to the Zeeman splittings of the two DQDs matching the resonator's frequency: $B_{1} = B_{2} = \omega_r$. We also note that in our numerical simulations we model up to six photons per resonator mode  -- this guarantees a very good approximation of the exact dynamics.

\begin{figure}[tb]
	\begin{centering}
	    \includegraphics[width=0.48\textwidth]{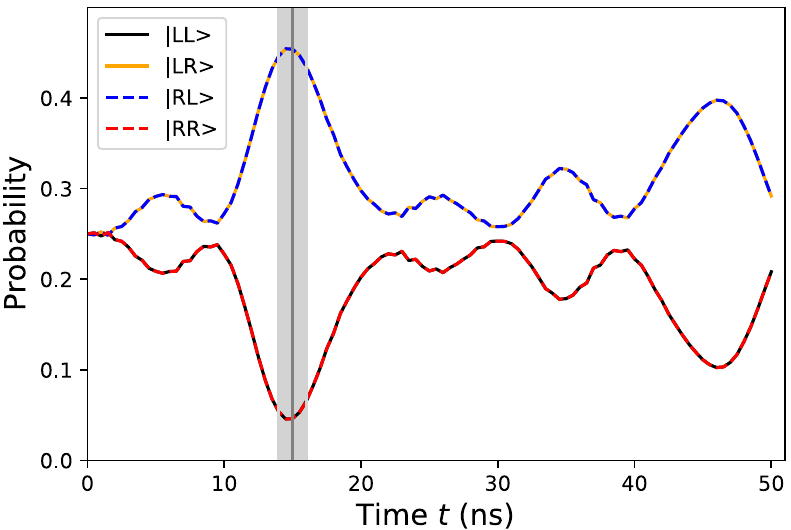}\\
	    \includegraphics[width=0.48\textwidth]{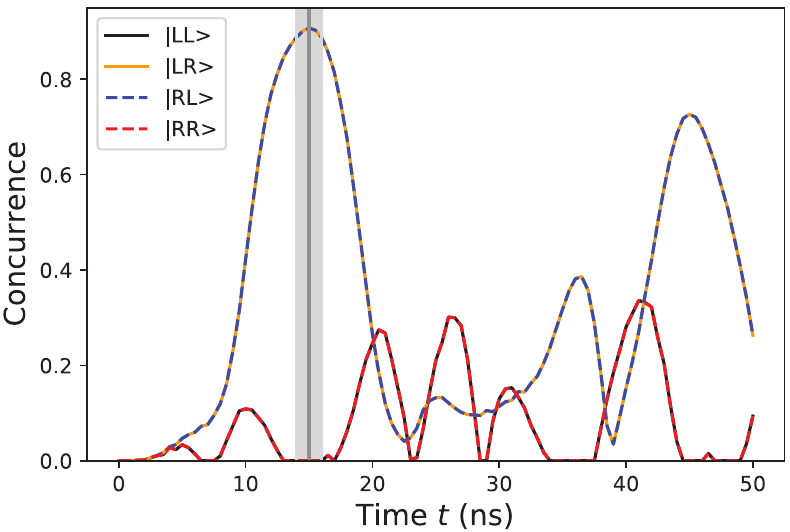}
		\caption{
			(a) Evolution of the probabilities of measuring the charges at different locations.
			(b) Evolution of the concurrence between the distant spins. Both plots were computed using the optimised evolution parameters. The optimal initial state is given by: $\ket{\psi} \approx \frac{\left(\ket{\text{L}}-\ket{\text{R}}\right)\ket{\uparrow}}{\sqrt{2}} \otimes\frac{\left(\ket{\text{L}}-\ket{\text{R}}\right)\ket{\uparrow}}{\sqrt{2}}\otimes\ket{0} $, with the first two terms describing the state of the first and second electrons respectively, and the last one the state of the cavity. 
			We account for potential difficulties to stop the interaction exactly at the maximal probability: we averaged the final density matrix over different stopping times (shaded area) weighted by a Gaussian distribution centred at $\SI{15}{ns}$ (see text).
			Here, $T_{2,s} = $~\SI{120}{\micro\second} and $ T_{2,c} = 400$~\SI{}{\nano\second}.
			\label{fig:prob_concurrence}
		}
	\end{centering}
\end{figure}

\subsection{Bell pair generation\label{sec:bell_gen}}
 In this section, we explicitly simulate the above introduced model and numerically search for a set of optimal evolution parameters. We explicitly take into account hardware noise that is comparable to that of current devices. We find that our cavity-based system is capable of generating raw Bell pairs of sufficiently high fidelity that we can then use as input for purification.

\begin{figure*}[tb]
	\begin{centering}
		\includegraphics[width=\textwidth]{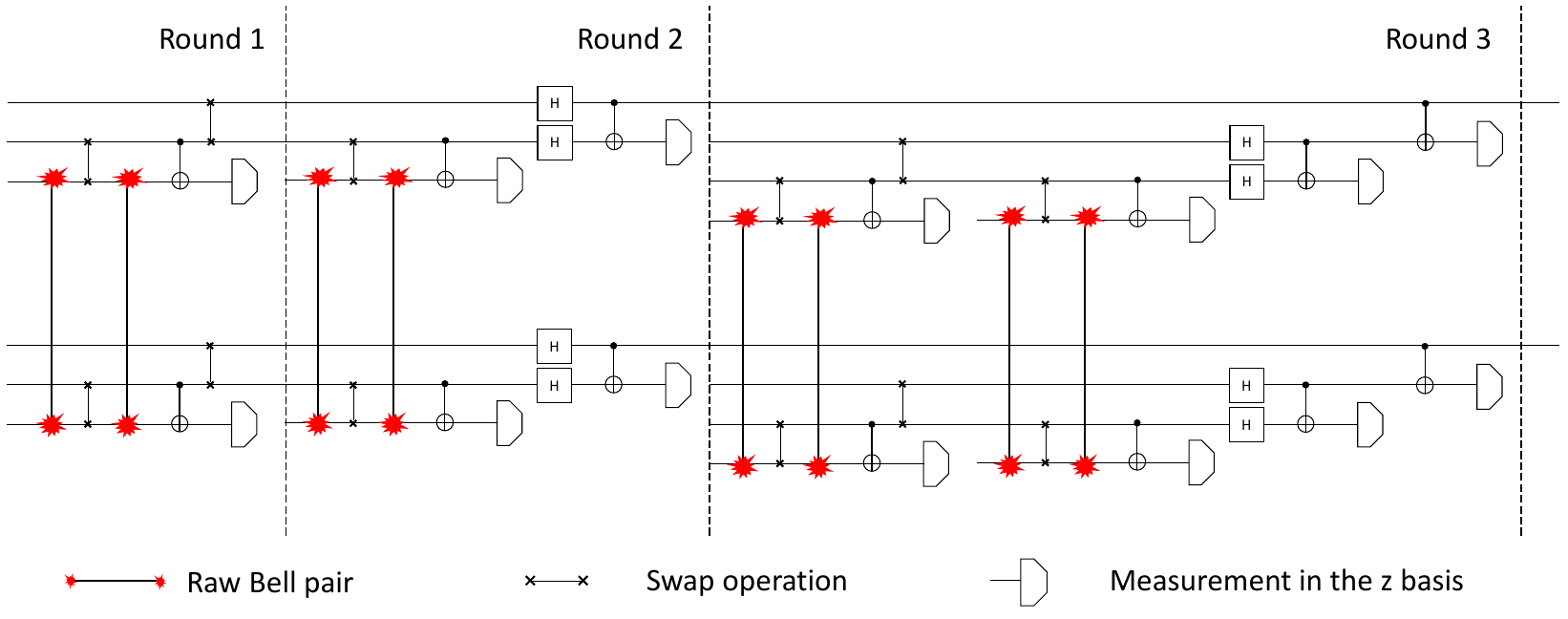}
		\caption{Entanglement purification protocol circuit. Here three rounds are displayed but depending on the noise model two rounds might be sufficient. The first round addresses bit flip errors in the input noisy, raw Bell pairs while the second round reduces phase flip errors. Then, in the third round we repeat the same procedure as in the first round but with two level-2 Bell pairs as input. The fourth round (not shown here), reuses the circuit in round 2 but with two level 3 Bell pairs as input. 
		If both bit flip and phase flip errors are present, the protocol should terminate at either the second or the fourth round depending on the noise rate. Errors in the purification process are accounted for by adding depolarizing noise after each local one and two-qubit gate, and measurement errors ($99.9\%$ fidelity) at each round (not displayed here). While the figure shows the canonical choice in entanglement purification, we tailored the protocol towards the specific error model of cavities by inserting further single-qubit rotations.
		\label{fig:distillation_circuit}
		}
	\end{centering}
\end{figure*}

In particular, we numerically simulate the evolution of the system's state under the Hamiltonian defined in \cref{eq:two_qubit_hamiltonian} while also accounting for different noise processes via the general Lindblad master equation as
\begin{align}
    \frac{\partial \rho}{\partial t} =  - i \left[H_{srs},\rho\right] {+} \sum_{i}\gamma_{i}\left(A_i\rho A_i^{\dagger}{-}\frac{1}{2}\left\{ A_{i}^{\dagger}A_{i},\rho\right\}\right).
    \label{eq:lindbladian_cavity}
\end{align}
Here $\gamma_i$ are positive rates, and $\hat{A}_{i}$ are Lindblad terms. Thereafter we denote each noise process using a tuple $(\gamma_i,A_i)$. In this work, we focus on three sources of noise.
First, we consider cavity loss $(\kappa,a)$ as defined by a cavity loss rate $\kappa$ and the annihilation operator $a$. Then, we consider spin dephasing $\left(1/(2T_{2,s}),s_z\right)$ characterised by the spin decoherence time $T_{2,s}$ and the Pauli operator $s_z$. Finally, we consider charge noise $\left(1/(2T_{2,c}),\tau_z\right)$ as defined by the charge decoherence time $T_{2,c}$ and the Pauli operator $\tau_z$. Our model assumes a time-independent Hamiltonian and the high-frequency fluctuations of the detunings are accounted for in the charge noise model \cite{abadillo-urielLongrangeTwohybridqubitGates2021a}. While this charge dephasing $T_{2,c} \leq 400$ ns is significantly faster than
spin dephasing, we note that in our protocol quantum information is never stored in the charge state -- the charge degree of freedom
merely acts as an interaction mediator for a brief period of time $\approx 15$ ns.

As described above, our goal is to entangle the two remote spins by generating a specific Bell state -- which may be noisy due to imperfections. We stress that our model in \cref{eq:two_qubit_hamiltonian} implemented in the resonant regime leads to highly non-trivial evolutions and even with no decoherence it cannot perfectly generate a Bell pair as detailed in \cref{app:state_analysis}. 
Intuitively, the reason is that we assume we only have the ability to abruptly alter
the interaction Hamiltonian --
and in the full Hamiltonian we only have control of at most $10$ individual parameters, whereas at the end of the evolution
a number of restrictive conditions need to be satisfied. First, the two spins have attained the highest degree of entanglement; second, the spins are separable from the charge degrees of freedom and from the cavity; third, the cavity is nearly in the vacuum state. As such, given the dimensionality of the Hilbert space is $2^4 \times N_{cav}$ (allowing for the fact that only the first $N_{cav}\approx 7$ energy levels are populated in the cavity), our dynamical system is clearly under-parametrised.
Therefore we optimise the system's trajectory to come as close as possible to the desired condition, but do not expect to perfectly meet it.

In order to obtain the best possible fidelity, we will perform projective measurements at a given optimal time on the interaction-mediating charge degrees of freedom, thereby also increasing the entanglement in our raw Bell pairs at the cost of a slight probability of failure as explained in \cref{app:state_analysis}. Our ultimate goal is that given a sufficiently high level of entanglement of our raw Bell pairs, we can improve its fidelity to arbitrarily high levels by using purification techniques.

To find an optimal set of parameters in \cref{eq:two_qubit_hamiltonian}, we performed a gradient-based optimisation that maximises the entanglement between the spins. The optimal parameter values can be found in \cref{app:parameter_regime}.
As a figure of merit for the optimisation, we use the concurrence \cite{woottersEntanglementFormationArbitrary1998} of the joint state of the two spins given a fixed charge measurement outcome. 
Averaging this concurrence for the charge measurement outcomes that we accept then defines our cost function. 
We also introduce a set of constraints for the optimisation. Firstly, the resonant condition $B_{1} = B_{2} = B = \omega_r$, and, secondly $2t_{c,1} = 2t_{c,2} = 2t_c = B$ (achievable by tuning the tunnelling barrier of the DQD \cite{martinsNoiseSuppressionUsing2016}) which gives us the maximum coupling between the spin and the charge \cite{benitoInputoutputTheorySpinphoton2017}. Apart from these fixed parameters, we optimise all other parameters that define our model and the initial state of each electron. It is interesting to reflect that, given a real device, one could perform a similar optimisation procedure (e.g. post fabrication) to find the device's ideal mode of operation.

In \cref{fig:prob_concurrence}, we report the evolution of our experimental system under an optimal set of parameters. In particular, \cref{fig:prob_concurrence}(a) shows the probabilities of different charge measurement outcomes, while \cref{fig:prob_concurrence}(b) shows the corresponding evolution of the concurrence between the spins. The concurrence peak nicely matches a probability peak given charge measurement outcomes in the odd parity subspace (as defined by the states $\ket{\text{LR}}$ and $\ket{\text{RL}}$).
We discuss in \cref{app:state_analysis} that the discarded charge measurement outcomes eliminate the
dominant coherent error in which the charge state $(\ket{\text{LL}}{-}\ket{\text{RR}})/\sqrt{2}$ is coupled to the unentangled spin state
$\ket{\downarrow \downarrow}$ due to energy conservation on-resonance -- which explains the nearly zero concurrence of the discarded measurements in \cref{fig:prob_concurrence}(b)(black and red lines).

\begin{figure}
	\centering
	\includegraphics[width=0.48\textwidth]{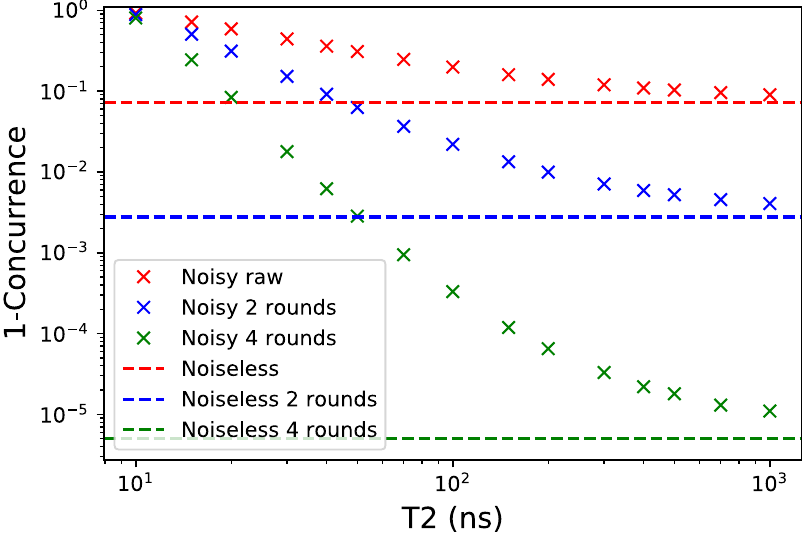}\\
	\includegraphics[width=0.48\textwidth]{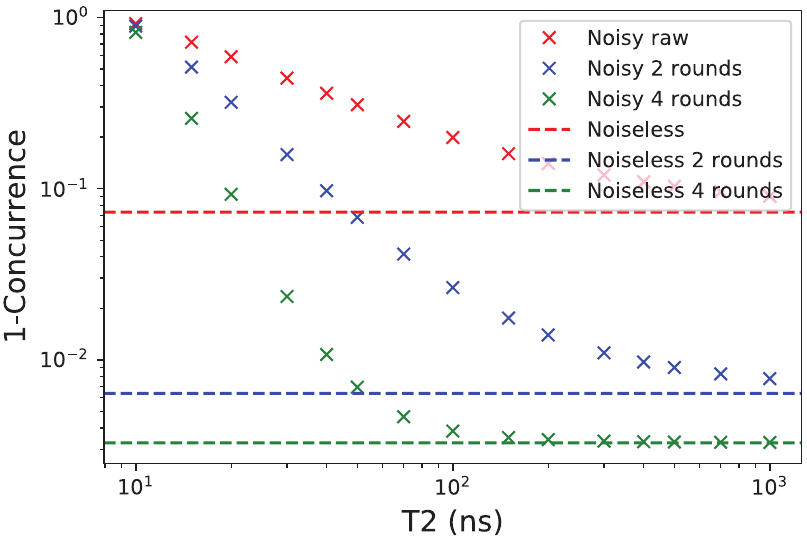}
	\caption{Concurrence (measure of entanglement) as a function of charge decoherence for different levels of purification. Above (below) without (with) errors in the local quantum gates in the purification process.
		The dashed Red line represents a perfect unitary evolution (no decoherence) and perfect charge measurements  and the only source of imperfections is the fact that our fast, resonant dynamics in \cref{eq:two_qubit_hamiltonian} cannot exactly generate a Bell pair.}
	\label{fig:purification_four_phase}
\end{figure}

As we perform a charge measurement at the maximal peak, i.e., at around $t \approx 15$~\SI{}{\nano \second}, we find a high success probability $\approx 0.9$ while the resulting state achieves a relatively high concurrence $\approx 0.89$.
In our simulations we also took into account the possible uncertainty in stopping the interaction instantaneously  by averaging over a Gaussian distribution of different stopping times with $\sigma = \SI{0.5}{ns}$ (grey shaded area in \cref{fig:prob_concurrence}).
Given state-of-the-art technology achieves smaller jitter by orders of magnitude, ours is a very conservative estimate to account for the possibility of integrating (likely lower quality) pulse-control electronics within the chip. 
Furthermore, the purified fidelities we observe are nearly oblivious to jitter times, in particular we observe fidelities
$99.60\%$,
$99.59\%$,
$99.51\%$
with a jitter time of
$0$ ps,
$50$ ps,
$500$ ps,
respectively,
which speaks for the robustness of our approach.
It is also worth noting that as we accept measurement outcomes in the odd parity subspace, there is a high probability ($\approx 0.96$) to find the photon in the vacuum state $\ket{0}$ (see \cref{sec:photon_after_meas}). This guarantees with high probability that we can promptly use the cavity in a next instance for generating another Bell pair.

\begin{table}[tb]
	\begin{ruledtabular}
		\begin{tabular}{cccc}
			$T_{2,c}$~(\SI{}{\nano \second}) & 400 & 100 & 50 \\ 
			\hline
			\\[-2mm]
			Fidelity & 94.5\% & 90.0\% & 84.4\% \\
			\\[-3mm]
		\end{tabular}
	\end{ruledtabular}
	\caption{\label{tab:pre_purification_fidelities} Pre-purification fidelities for different noise parameters. The fidelities are expressed with respect to the Bell pair $\ket{\psi^{-}} = \frac{1}{\sqrt{2}}\left(\ket{01}-\ket{10}\right)$.} 
\end{table}

It is expected that in typical experiments the coupling between each spin and the cavity is electrically controllable \cite{miCoherentSpinPhoton2018,benitoInputoutputTheorySpinphoton2017}. This allows us to rapidly switch off their interactions at our optimal evolution time---which corresponds to the peaks in \cref{fig:prob_concurrence}---by electrically adjusting either detuning or the tunnelling couplings. However, it might be difficult to stop the interaction exactly at the maximal probability in practice. We take this into account by averaging the final density matrix over different stopping times weighted by a Gaussian distribution centred at $\SI{15}{ns}$ (with $\sigma = \SI{0.5}{ns}$). As an artefact, the present dynamics might introduce a deterministic phase to our quantum state, however, if this phase is stable over consecutive generations of Bell pairs it is automatically removed by our distillation process.

We also note that the quality of the entanglement in the raw Bell pairs (without purification) might not be sufficient to perform high-fidelity long-range operations as we report in \cref{tab:pre_purification_fidelities}. In particular, we expect that charge dephasing noise (which deteriorates our hybridised states) is the dominant source of error and we report fidelities in \cref{tab:pre_purification_fidelities} for different values of $T_{2,c}$ decoherence times on the order of tens up to hundreds of nanoseconds. For each simulation we assumed a spin decoherence time of $T_{2,s} = $~\SI{120}{\micro\second}. An ensemble-averaged decoherence time $T^{*}_{2,s} =$~\SI{120}{\micro\second} has been achieved experimentally \cite{veldhorstAddressableQuantumDot2014}, hence we expect our $T_{2,s}$ to be at least equal. It is worth noting that $T_{2,s}$ is not a bottleneck,
and while it could be reduced without significantly affecting our results, it is expected that $T_{2,s}$ will
have to be reasonably large given for any practical application one needs to perform at least hundreds of quantum gates.

\begin{table}[tb]
	\begin{ruledtabular}
		\begin{tabular}{cccc}
			$T_{2,c}$~(\SI{}{\nano \second}) & 400 & 100 & 50 \\ \hline
			\\[-2mm]
			2 rounds & 99.5\% (99.7\%)& 98.7\% (98.9\%)& 96.6\% (96.8\%)\\
			\\
			4 rounds & 99.8\% ($>$99.9\%)& 99.8\% ($>$99.9\%)& 99.7\% (99.9\%)\\
			\\[-3mm]
		\end{tabular}
	\end{ruledtabular}
	\caption{\label{tab:post_purification_fidelities}
	Post-purification fidelities for different noise parameters and rounds of purification.
	In parentheses: without noise in the local operations in the
	purification process. The fidelities are expressed with respect to $\ket{\phi^{+}}$.
}
\end{table}
\subsection{Entanglement purification}
As noted in the introduction, as early as 2003 entanglement purification was explored theoretically as a means to distribute a quantum task over subsystems with non-optimal links~\cite{durEntanglementPurificationQuantum2003,krastanovOptimizedEntanglementPurification2019,nickersonFreelyScalableQuantum2014,lopiparoResourceReductionDistributed2020,nigmatullinMinimallyComplexIon2016}.
We briefly recapitulate the basic two-node version entanglement purification (or, `distillation' --  the terms are now typically used interchangeably). The `Alice' and `Bob' nodes require a means to create Bell pairs between them, as well as a modest set of local operations. Purification involves stages, or rounds, where two Bell pairs are locally entangled and then one pair is measured out. Alice and Bob compare their measurement outcomes over a classical channel. If the Bell pairs were both noise-free then only certain measurement outcomes are possible; if the allowed outcome is not seen then Alice and Bob discard the remaining Bell pair, otherwise they store it. Thus, a surviving Bell pair has passed a form of validation and its fidelity is higher: in the circuits used here, one noise channel (say, phase) will have been suppressed from order $p$ to $p^2$ while the other (flip) increases from $p$ to $2p$. Thus, one must create a second validated Bell pair, and then combine the two pairs in a new, higher round of purification that targets the remaining noise channel. In this way, four `raw' Bell pairs with infidelity of order $p$ can be combined to give rise to a single pair with infidelity of order $p^2$ (with a modest prefactor). Of course, further rounds can be employed to suppress noise to order $p^3$ or higher, but the cost in Bell pairs increases exponentially.

Noise in the local operations is relevant but the purification process itself combats it, so that only noise in the final round is significantly impactful. Importantly, if there is significant structure in the noise in the \emph{raw Bell pairs}, then the purification process can be tailored to exploit this structure -- structureless full rank noise (white noise) is the least desirable.

We use the controlled interaction between the spin qubits and the cavity described in the previous section as a process to generate noisy bell pairs.
In \cref{fig:purification_four_phase} we plot the concurrence (entanglement) in the Bell pairs after various rounds of
purification as a function of the dominant noise source, the charge decoherence time $T_{2,c}$ -- while we also model spin decoherence
and imperfections in the local single and two-qubit gates via depolarizing noise as well as measurement errors. Here, we use the variant of the standard purification circuit in \cref{fig:distillation_circuit} that gives us the best concurrence after the purification process for the particular structure of noise in our raw Bell pairs. Furthermore, in \cref{app:alt_purification} we report that by adding further single-qubit gates we can obtain better concurrence in the ideal purification case.
For simplicity, in the following we will mostly focus on the $T_{2,s} =$~\SI{120}{\micro\second} and $ T_{2,c} = 400 $~\SI{}{\nano\second} case. Such a charge decoherence rate can be reached with current hardware \cite{miStrongCouplingSingle2017}.

\begin{figure}[tb]
	\begin{centering}
		\includegraphics[width=0.45\textwidth]{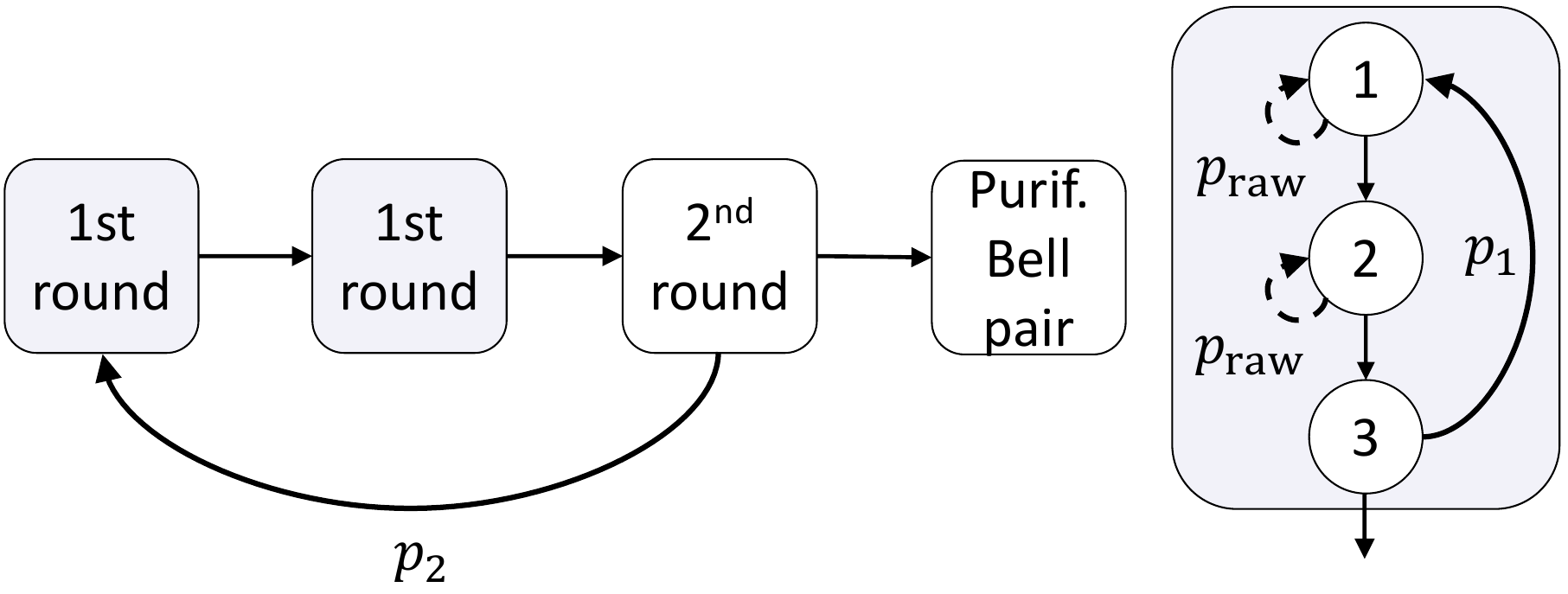}
		\caption{
			Distillation process flow chart for two rounds of distillation. The blue-shaded box represents the steps for one round of purification. It requires the generation of two raw Bell pairs (steps 1 and 2), a process that fails with probability $p_{\text{raw}} \approx 0.10$. The distillation process (step 3) fails with probability $p_{1} \approx  0.08$. After having successfully generated two level-$1$ Bell pairs, we perform a second round of distillation to get the final purified Bell pair, which fails with probability $p_2 \approx 0.06$. The probabilities are given for the $T_{2,s} =$~\SI{120}{\micro\second} and $ T_{2,c} = 400$~\SI{}{\nano\second} case. 
			\label{fig:flow_chart}
		}
	\end{centering}
\end{figure}

Since in entanglement purification we need to discard certain measurement patterns, it is a non-deterministic process. In our case, the purification protocol (in both rounds) succeeds if we measure both qubits in the same state. \cref{fig:flow_chart} shows the main steps of the distillation process. On average, one needs to generate $N_{avg} \approx 5.2$ noisy Bell pairs to obtain a $99.5 \%$ fidelity Bell pair. 
Assuming typical gate/measurement times this results in an average generation rate $g_{Bell} \approx 0.14$ MHz (see \cref{app:generation_rate}),
even though in the limit of very fast gates and measurements
entanglement generation is dominated by the cavity interaction time with a 
generation rate of $g_{Bell} \approx 12.9$ MHz.

Fortunately this rate of $0.14$ MHZ can still suffice for powerful applications including the derangement-based error mitigation described in \cref{sec:applications}. Prior experiments demonstrated that coherence can be
maintained with respect to phase noise for as long as $28$ ms using a Carr-Purcell-Meiboom-Gil sequence \cite{veldhorstAddressableQuantumDot2014}, and indeed even devices fabricated in $300$ mm commercial wafers have achieved a $T_2$ time of $3.7$ ms \cite{zwerverQubitsMadeAdvanced2022}.
We therefore expect our approach allows the generation of tens or even hundreds of purified Bell pairs to be used for `comparing qubits' between cores using the derangement method in \cref{sec:applications} before dephasing would negate the advantage of the process.

For applications where this rate is insufficient, note that the cavity is in use for only a very small fraction of the procedure ($\approx 15$ ns), and we can thus
significantly improve generation rates by a `parallelisation' of the purification step: The extended architecture would include
$n$ `purification stations' on both sides of the cavity, and while the cavity is exclusively used to generate a single Bell
pair at a time, these raw Bell pairs are fed in an alternating way into the different purification stations.
By choosing a modest parallelisation $n \leq 5$ we expect the generation rate is effectively increased by the same factor
$n$.
We also remark that our particular purification protocol transforms the initial $\ket{\psi^{-}}$ into $\ket{\phi^{+}}$.

For charge noise worse than $ T_{2,c} = 400~\SI{}{\nano\second}$, four rounds of purification are necessary to obtain an adequate fidelity. In \cref{tab:post_purification_fidelities} we collect the fidelities for different charge noise parameters for two and four rounds of purification. The fidelities reported in parentheses are obtained without errors in the local gates and during measurements in the purification process. Let us define the obtained Bell pair after the $i^{th}$ successful purification round as a Bell pair of level $i$. Four rounds of distillation is rather demanding, since in order to go from the $i^{th}$ round to the $(i{+}1)^{th}$ one needs two level-$i$ Bell pairs. As purification is a stochastic process, performing more rounds leads to a decrease in the success probability and hence to a smaller generation rate.
See \cref{app:generation_rate} for its computation for $T_{2,c} = 100 ~\SI{}{\nano \second}$ and $T_{2,c} = 50 ~\SI{}{\nano \second}$.

With better hardware, the number of purification rounds will decrease while the generation rate and the fidelity will increase. Indeed, with negligible decoherence, one obtains a fidelity
$\approx 99.9\%$ with $g_{Bell} \approx 0.15$ MHz after two purification rounds 
($g_{Bell} \approx 13.9$ MHz via fast gates and measurements). However, we will see in \cref{sec:applications} that for many applications one does not need to generate perfect Bell pairs and already a fidelity of $99.5\%$ is enabling for powerful NISQ applications.

\section{Shuttling}

Perhaps the most straightforward means of transporting quantum information on-chip is through the direct movement of the electrons themselves through a series of quantum dots. So-called shuttling has featured in many spin-based quantum computing architecture proposals as an efficient means of transporting spins over micron-scale distances without creating a time-bottleneck \cite{taylorFaulttolerantArchitectureQuantum2005,boterSparseSpinQubit2019,buonacorsiNetworkArchitectureTopological2019,jadotDistantSpinEntanglement2021}. In the following we focus on a ``bucket-brigade" shuttling protocol which is perhaps the most theoretically well-understood alternative in silicon, in contrast to ``conveyor mode" or ``surface-acoustic-waves" protocols.

The unit operation of a shuttling protocol is the coherent tunnelling of an electron from one quantum dot to the next. This may be actuated via a time-dependent electrostatic detuning $\epsilon(t)$ present between left (L) and right (R) dots, such that the charge state is adiabatically tipped into the target dot. The velocity of the detuning sweep must be sufficiently slow to minimise diabatic transitions to excited valley-orbit states, while slow enough to ensure charge noise does not induce transitions near avoided crossings. Such restrictions are dependent on device design and the degree of pulse optimisation, though speeds on the order of 1 nanosecond per tunnelling event are considered feasible \cite{buonacorsiSimulatedCoherentElectron2020,krzywdaAdiabaticElectronCharge2020,krzywdaInterplayChargeNoise2021}. With a dot-to-dot spacing of 50-\SI{100}{\nano\meter}, the distance of a few microns can be traversed in tens to hundreds of nanoseconds, comparable to single- and two-qubit gate times in silicon quantum dot processors \cite{watsonProgrammableTwoqubitQuantum2018}.

Coherent spin shuttling has been realised over effective several-micrometer distances in a GaAs quantum dot circuit \cite{flentjeCoherentLongdistanceDisplacement2017}. Reliable charge shuttling has also been shown in multi-dot Si/SiGe arrays \cite{millsShuttlingSingleCharge2019,seidlerConveyormodeSingleelectronShuttling2022}, while repeated coherent spin tunnelling between two Si-MOS dots has also been demonstrated \cite{yonedaCoherentSpinQubit2021}. This places shuttling as a top candidate for micron-scale on-chip quantum information transport in near-term devices.

Previous theoretical treatments have given substantial attention to the ability of a single shuttling event to preserve an arbitrary input state's fidelity \cite{huangSpinQubitRelaxation2013, liIntrinsicErrorsTransporting2017,zhaoCoherentElectronTransport2019,buonacorsiSimulatedCoherentElectron2020,ginzelSpinShuttlingSilicon2020}. However, a shuttling channel does not need to be able to shuttle arbitrary states to constitute a useful resource. In accord with the theme of the present paper, shuttling along arrays of quantum dots can provide a means of on-chip entanglement distribution and may be a component of gate teleportation or error correction schemes between many computational cores. In many ways, this is a less demanding task, as unitary transformations induced by the shuttling channel can be mitigated through calibration or distillation protocols. Rather than risk losing information in a shuttled data qubit, entanglement may be continuously generated on a timescale similar to native physical gates and then be used by spin registers separated by micron-scale distances on-demand. 

We present a schematic of two spin registers connected via a quantum dot shuttling channel in \cref{fig:ill}(c): A Bell pair consisting of two electrons (small orange dots) may be created using a high-fidelity two-qubit interaction near one spin register, made possible with a local micromagnet. One electron of the pair may then be shuttled down the few micron-length chain on a \SI{100}{\nano\second} timescale. A measurement dot (M) at the very end of the chain may be used to probabilistically project the electron's state into the ground valley. The resulting spin states of the Bell pair may be stored in the ancilla dots (A) via a SWAP interaction or state-preserving tunnelling. A second Bell pair may be initiated and transported in the same way. A single-round purification scheme may be run on the two Bell pairs, using local two-qubit interactions. Charge measurement can again be carried out using the ancilla dots, and a successful outcome yields a high-fidelity Bell pair on the memory qubit. This entangled pair may be used as part of an algorithm requiring interaction between the two spin registers.

To investigate the dynamics of shuttling, it suffices to focus on the process of shuttling an electron from one QD to another in a DQD structure~\cite{zhaoCoherentElectronTransport2019,ginzelSpinShuttlingSilicon2020}. The Hamiltonian we have is similar to the static DQD Hamiltonian in \cref{eqn:DQD_ham}, but now we are equipped with the ability to change the detuning between the two dots ($\epsilon \rightarrow \epsilon(t)$) for carrying out the shuttling. Without the strong magnetic field gradient induced by the micromagnet in the cavity-mediated interaction in \cref{sec:cavity_mediated_gate}, it becomes essential to take into account small magnetic fields. We consider the intrinsic spin-orbit interaction as the Rashba and Dresselhaus effects \cite{hansonSpinsFewelectronQuantum2007} which we model via the Hamiltonian
\begin{align*}
    H_{\mathrm{SOI}} = \tau_y \otimes (\,  \vec{\Omega}\cdot\vec{s}\,).
\end{align*}
Here $\tau_y$ and $s_i$ are Pauli matrices as explained below \cref{eqn:DQD_ham} in an effective magnetic field $\vec{\Omega}$ \cite{hansonSpinsFewelectronQuantum2007}.

As opposed to our modelling of the cavity-based distribution in \cref{sec:cavity_mediated_gate}, here we need to take into account the coupling with the valley degree of freedom which we identify as potentially the leading source of error in shuttling.
In particular, the time-dependent variation (sweeping) of the detuning leads to the possibility of coupling the charge qubit to higher excited states of the valley degree of freedom which we model as an effective \emph{valley qubit}
described by the two basis states as the bulk valley states $\ket{z}$ and $\ket{\bar{z}}$.
By using $v_i$ to denote the Pauli matrices $\sigma_i$ acting on the valley qubits, we define $v_{+} = \frac{1}{2}(v_x + iv_y)$.
Furthermore, we define the projectors that act on the charge degree of freedom as $\tau_{d} = |d \rangle \langle d |$ with $d \in \{L, R\}$ and thus we can write the valley-charge coupling term as
\begin{align}\label{eqn:valley_charge_coupling}
    H_{\tau v} = \sum_{d \in \{L, R\}} (\Delta_d \, \tau_d \otimes v_+  + \textrm{H.c}).
\end{align}
Here we denote the site-dependent valley-couplings $\Delta_d=|\Delta_d| e^{-i\phi_d}$ corresponding to the valley splitting energies $E_{V,d} = 2|\Delta_d|$, refer to Ref.~\cite{culcerInterfaceRoughnessValleyorbit2010} for more details.

\begin{figure}[tb]
\centering
\includegraphics[width=0.48\textwidth]{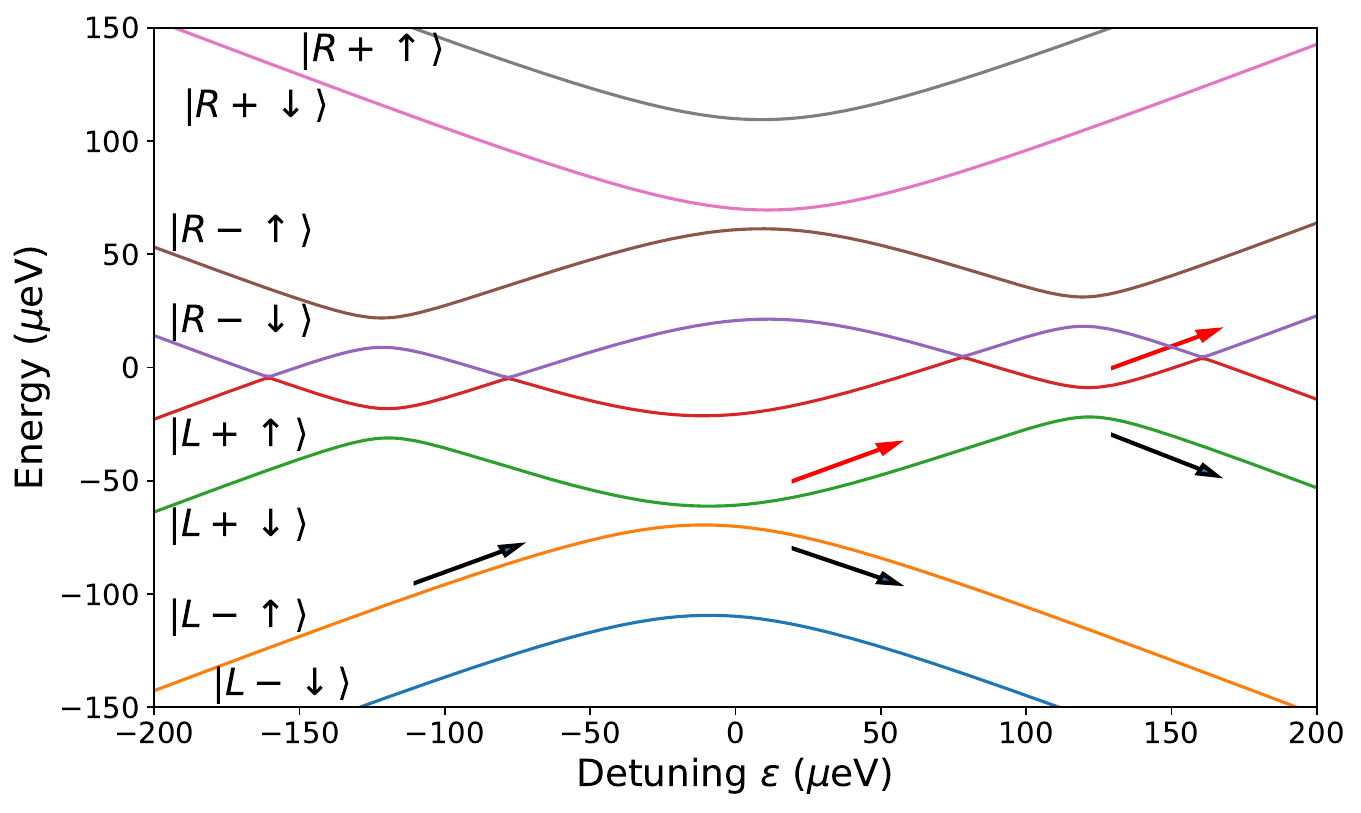}
\caption{
Energy level diagram of a silicon DQD containing spin, valley, and orbital degrees of freedom for $t_c=$~\SI{25}{\micro\eV}, $B=$~\SI{40}{\micro\eV}, $b_x=b_z=$~\SI{1}{\micro\eV}, $\phi_L = \pi/6$, $\phi_R=0$, $|\Delta_L|=$~\SI{60}{\micro\eV}, and $|\Delta_L|=$~\SI{70}{\micro\eV}. As the detuning is swept, an electron state traverses from left to right. Both adiabatic (black arrows) and diabatic (red arrows) transitions may occur, and the initial spin state will coherently evolve into a superposition of output states. Leakage to the excited orbital sector will result in a mixed spin-valley state.}
\label{fig:shuttling_1}
\end{figure}

Hence, the full Hamiltonian that we will consider for the shuttling process is:
\begin{align}\label{eqn:H_Si_DQD_global}
    H_{\mathrm{sh}} &= H_{\mathrm{DQD}}(t) + H_{\mathrm{SOI}} + H_{\tau v}
\end{align}
where $H_{\mathrm{DQD}}(t)$ is \cref{eqn:DQD_ham} up to the time-dependent detuning $\epsilon \rightarrow \epsilon(t)$. 

Due to the local magnetic field variance, the local natural spin basis is different from the global basis defined by $B$ in \cref{eqn:DQD_ham}, and we can transform \cref{eqn:DQD_ham} into the local spin basis via a unitary transformation defined in Ref.~\cite{ginzelSpinShuttlingSilicon2020}. Similarly, the local valley eigenstates are also different from the bulk valley states $\{\ket{z},\ket{\bar{z}}\}$. Suppose the valley eigenstate at site $d\in\{L,R\}$ are $\ket{-}_d$ (ground state) and $\ket{+}_d$ (excited state), we can then express them as
\begin{equation}
    \label{eqn:valley_eigenstates}
    \ket{\pm}_d = \frac{1}{\sqrt{2}}(\ket{z}\pm e^{i\phi_d}\ket{\bar{z}}).
\end{equation}
When the valley-orbit sector of \cref{eqn:valley_charge_coupling} is rewritten in the $\{\ket{\pm}_d\}$ basis, the valley-conserving and valley-flipping tunnel couplings can be identified specifically as \cite{zhaoCoherentElectronTransport2019}
\begin{align}
    \label{eqn:tvc}
    t_{vc} = \frac{t_c}{2}(1+e^{-i\delta\phi}),\quad t_{vf} = \frac{t_c}{2}(1-e^{-i\delta\phi}).
\end{align}
Here $\delta\phi = \phi_L - \phi_R$ is the difference in phases between the two valley coupling parameters $\Delta_L$ and $\Delta_R$. This phase difference depends precisely on the overlap of the electron wavefunction with the quantum well interface \cite{culcerInterfaceRoughnessValleyorbit2010}. When $\delta\phi = 0$, states of opposite valleys cross entirely, and when $\delta\phi=\pi$, valley flips occur deterministically regardless of the magnitude of the bare tunnel coupling, making this microscopic parameter crucial in the description of electron shuttling in silicon.

\cref{fig:shuttling_1} illustrates the spin-valley-orbit energy level diagram of $H'$ for a realistic set of parameters that may be encountered during shuttling as detailed in \cref{sec:shuttling}. Initially, an electron will populate the lowest two states $\ket{L - \downarrow}$  and $\ket{L-\uparrow}$. Population in the former ground spin state will adiabatically transfer reliably to $\ket{R-\downarrow}$, as the unique avoided crossing near zero detuning is large. Population in the latter excited state encounters a much smaller effective tunnel coupling $t_{vc}$ near zero detuning, possibly allowing diabatic Landau-Zener transitions into the higher energy level. Subsequent avoided crossings provide further opportunities for population leakage into excited spin-valley-orbit states.

If the quality of the shuttled state is measured on the basis of fidelity, population loss into higher-energy states will degrade the quality of our Bell pairs unless recovered through appropriate calibrations \cite{zhaoCoherentElectronTransport2019,buonacorsiSimulatedCoherentElectron2020}, i.e., such a unitary evolution does not inherently destroy information but does rather introduce a coherent error.
Although the coherent error introduced by the unitary evolution under our Hamiltonian in \cref{eqn:H_Si_DQD_global}
is local to the shuttled spin, it may increase the spin-valley entanglement and thus lead to a decreased entanglement between 
the spin Bell pairs due to the monogamy of entanglement.

Let us now identify two additional mechanisms of central importance for information loss while a further discussion of our assumptions and the impact of other noise sources is included in \cref{sec:shuttling}.

First, at the end of the detuning sweep, the state may populate the excited orbital state $\ket{L}$, corresponding to the charge ``boucing back" to the initial dot. In any practical shuttling implementation, the detuning will be plunged much further, such that this excited orbital state will inevitably cross with higher energy orbital states in the target dot. For a sufficiently large detuning, charge transfer will occur deterministically but can populate these excited orbits which are not included in \cref{eqn:H_Si_DQD_global}. Nevertheless, excited orbitals in silicon quantum dots relax on a sub-nanosecond timescale \cite{tahanRelaxationExcitedSpin2014}, and therefore the ground orbital state $\ket{R}$ of the target dot will be entirely occupied prior to a subsequent shuttling operation.

\begin{figure}[tb]
\centering\includegraphics[width=0.48\textwidth]{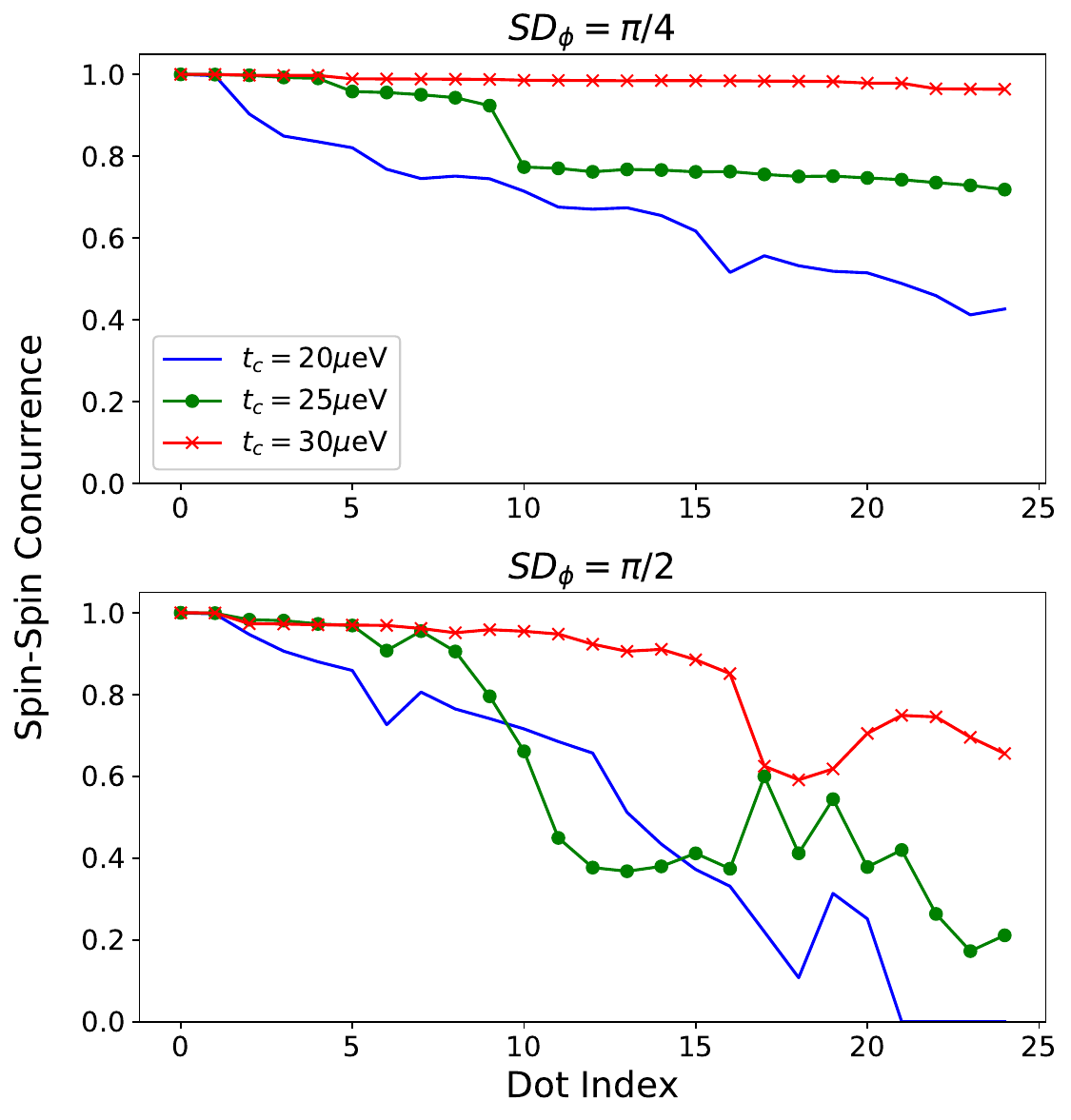}
\caption{Concurrence between two spins, initialised in a Bell state, as one is shuttled down an array of 25 quantum dots. The value associated with each dot $i$ corresponds to the concurrence within the ground valley subspace at the end of an $i$-quantum-dot chain. Valley splittings and magnetic fields are identical for all traces, while the effect of valley phase fluctuations and different bare tunnel couplings is studied. Valley phase differences are sampled from a normal distribution with mean 0 and (top) $\sigma =\pi/4$ while (bottom) $\sigma=\pi/2$.
Note that increasing (decreasing) entanglement between the valley degree of freedom and the shuttled spin decreases (increases) the concurrence between the (initially maximally entangled) two spins due to the monogamy of entanglement.
The plots show a typical instance of the effect of random phase differences as one would encounter in a given physical device, whereas the fidelity figures we show
later are averaged over a large number of random instances.}
\label{fig:shuttling_2}
\end{figure}

Second, a finite population of the excited valley state $\ket{+}_R$ may also lead to decoherence. Although the bare valley degree of freedom is believed to be long-lived, hybridisation with the orbital state may substantially decrease the coherence time as a result of quantum well interface roughness \cite{tahanRelaxationExcitedSpin2014,borossControlValleyDynamics2016}. Therefore, in a manner similar to the population of the excited orbital states, excited valley population and decoherence can cause overall information loss. However, the spin-valley state may plausibly be well-defined during the entire duration of a \SI{100}{\nano\second} shuttling operation. While spin-valley computation is universal \cite{rohlingUniversalQuantumComputing2012}, and coherent valley control has been demonstrated \cite{schoenfieldCoherentManipulationValley2017,penthornTwoaxisQuantumControl2019}, fault-tolerant fidelities have only been achieved with spins \cite{xueQuantumLogicSpin2022,noiriFastUniversalQuantum2022}. Therefore, among many creative possibilities, we find it prudent to projectively measure the valley state of the shuttled electron and post-select on ground state measurements. This leaves the spin state intact while introducing a slight probability of failure
as we may need to discard certain measurement outcomes. We outline how this may be accomplished in \cref{sec:shuttling}.

While it is the decoherence properties of the excited valley-orbit states that ultimately lead to information loss, it is the parameters of \cref{eqn:H_Si_DQD_global} that determine the extent to which these states become populated during shuttling. The electrostatic and magnetic environment of the shuttling chain can be accurately engineered, or at least known, to good accuracy. However, the variation in the valley parameters $\Delta_d$ is believed to be large on account of their sensitivity to microscopic interface details~\cite{zimmermanValleyPhaseVoltage2017}, and experiment is just beginning to probe the variation in inter- and intra-valley tunnel couplings in silicon quantum dot arrays~\cite{borjansProbingVariationIntervalley2021}.

In \cref{fig:shuttling_2}, we emphasise the paramount importance of valley phase uniformity for Bell pair distribution via shuttling. We consider chains of 25 quantum dots, with each chain having a uniform bare tunnel coupling $t_c$ between adjacent sites. A protocol as described in \cref{fig:shuttling_1} is run, while the spin-spin concurrence between a stationary and shuttled spin is evaluated down the chain as a measure of entanglement \cite{cavesEntanglementFormationArbitrary2001}, as if the shuttling protocol was halted at each location on separate experiments. For valley phase differences randomly selected from a Gaussian distribution with a modest standard deviation of $SD_\phi = \pi/4$, preserving high concurrence depends on adjacent dots having a sufficiently large bare tunnel coupling $t_c$. The threshold is roughly given by the condition for a vanishing anticrossing between $\ket{L-\uparrow}$ and $\ket{R-\downarrow}$ near zero detuning when $B\approx2|t_{vc}|$, corresponding to maximal mixing of the spin and valley states. For larger variation in the valley phase differences, the higher probability for spin-valley mixing with each tunnel coupling results in a faster decrease of spin-spin concurrence with chain length for all bare tunnel couplings.

In the Appendix, in \cref{fig:shuttling_charge_noise} we also show a representative Pauli Transfer Matrix (PTM) for shuttling chains with parameters that result in high spin-spin concurrences. From this PTM, the shuttling superoperator $\mathcal{S}$ can be interpreted as a coherent z-rotation as well as amplitude damping towards the $\ket{\downarrow}$ state as indicated by the nonzero $I_{in}\rightarrow Z_{out}$ element \cite{greenbaumIntroductionQuantumGate2015}. Such an effect can be understood as the result of the $\ket{\uparrow}$ state being principally involved in diabatic transitions during the tunnelling operations. As such, let us consider the action of the PTM on an initial Bell state $\rho = \ket{\psi_+}\bra{\psi_+}$ as

\begin{align}
    \label{eqn:purifiable_state}
    \rho' & \approx \ket{\psi_+'}\bra{\psi_+'}+\epsilon\ket{\downarrow\downarrow}\bra{\downarrow\downarrow} \\
          & = \begin{pmatrix} 0\ \ & 0 & 0 &\ 0\\
                            0\ \ & \frac{1}{2}-\epsilon & e^{i\phi}\sqrt{\frac{1}{2}(\frac{1}{2}-\epsilon)} &\ 0\\
                            0\ \ & e^{-i\phi}\sqrt{\frac{1}{2}(\frac{1}{2}-\epsilon)} & \frac{1}{2} &\ 0\\
                            0\ \ & 0 & 0 &\ \epsilon \end{pmatrix}, \nonumber
\end{align}

\noindent where $\ket{\psi_+'} = \sqrt{\frac{1}{2}-\epsilon}\ket{\uparrow\downarrow}+e^{-i\phi}\sqrt{\frac{1}{2}}\ket{\downarrow\uparrow}$ is an asymmetric Bell state which has accumulated some phase $\phi$. When an ideal single-round purification circuit is applied to two copies of $\rho'$, a ``11"~$\leftrightarrow\ket{\downarrow\downarrow}$ outcome occurs with probability $(1-\epsilon)^2(\frac{1}{2}-\epsilon)$ with the resulting entangled state being the ideal Bell pair $\ket{\psi_+}$. Entangled states distributed via shuttling are therefore highly amenable to purification provided concurrence is mostly preserved. Of course, other decoherence processes do manifest beyond those captured in \cref{eqn:purifiable_state}, such that even an ideal purification circuit does not completely restore the initial state. For example, over many instances of the 25-dot chain with $t_c=$~\SI{30}{\micro\eV} and $SD_\phi=\pi/4$, concurrences after a single ideal round of purification average 99.5\%. Here we do not report rigorous success probability estimates as in case of the cavity-based alternative given these highly depend on the valley phase differences of the particular device.

\begin{figure}[tb]
\centering
\includegraphics[width=0.48\textwidth]{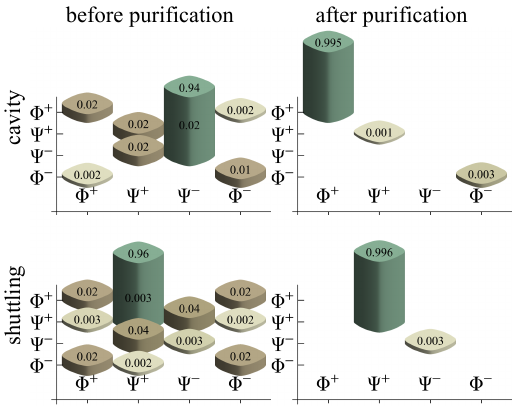}
\caption{
Comparing output density matrices in the case of cavity-based and shuttling-based
approaches. Absolute values of matrix elements in the Bell basis assuming imperfection parameters $T_{2,c}=\SI{400}{ns}$ (top) and $SD_\phi = \pi/4$ (bottom). The matrix associated to shuttling before purification corresponds to an average over multiple instances of 25-dot chains and is given up to local rotations. We need 2 rounds for the cavity-based approach, and only 1 for the shuttling-based one.
The largest fidelity with respect to the closest of the four canonical Bell states is increased from $94.5\%$ to $99.5\%$ for the cavity-based method, and from $95.7\%$ to $99.6\%$ for shuttling.
}
\label{fig:comparison}
\end{figure}

\section{Comparison of entanglement distribution modes}

Let us now compare the imperfect Bell pairs that we obtain via the cavity and suttling-based approaches. We show their corresponding density matrices before and after 2(1) rounds of entanglement purification in  \cref{fig:comparison}.

As such, we conclude that both mechanisms are able to provide sufficiently high grade shared Bell pairs, that subsequent entanglement purification yields final fidelities at about $99.5\%$ or better. Interestingly, the low-rank nature of the noise in the case of shuttling (only 2 brown bars in the diagonal of the density-matrix in \cref{fig:comparison} and not 3) has the consequence that the purification process can be more simple -- a single stage does suffice. The full-rank noise predicted for the cavity-mediated link (3 brown bars in the diagonal in \cref{fig:comparison}) does however require a multi-round purification to reach high grade final Bell states. A second distinction is that a shuttling channel could simultaneously transport a number of spins in a `pipeline' mode, whereas multiple simultaneous use of a cavity is possible in principle but may be more difficult to achieve in practice.

While these comparison points provide an interesting perspective, in reality it may be unlikely that a chip architect would select between these mechanisms; rather the choice would be determined by the desired length of the link. Shuttling is likely to be relevant in the $1-10$\,$\mu$m range, with cavity-based links appropriate for longer distances up to several mm (or even between chips).

\section{\label{sec:applications} Applications suited to Multicore}

Let us now consider a number of potential applications building on the above
interlinked multicore model.
The key characteristic here is that inter-core operations can be assumed to be of high fidelity but are a limited or `expensive' resource in comparison to `cheaper' intra-core operations which will be faster and (we suppose) capable of parallel operation over the core. We identify several important applications that are compatible with these features, and they may be of particular relevance to
near-term quantum quantum computers. We begin by arguing that recently proposed, exponentially powerful error mitigation techniques are eminently compatible with the multicore paradigm, and we simulate a VQE task enabled by such a mitigation.

\subsection{Exponential Error Suppression}

In the following we focus on the recently introduced Error Suppression by Derangements (ESD) and Virtual Distillation (VD) approaches \cite{koczorExponentialErrorSuppression2021,hugginsVirtualDistillationQuantum2021}. These error mitigation techniques can achieve exponential suppression of hardware noise, which is reminiscent of true quantum error correction. However, the technique requires significantly fewer resources than quantum error correction and is compatible with near term quantum devices. As such, we achieve exponential suppression by preparing $n$ identical copies of a computational quantum state, which fits well with our proposal of utilising multi-core quantum processors.

\begin{figure}[tb]
	\begin{centering}
		\includegraphics[width=0.45\textwidth]{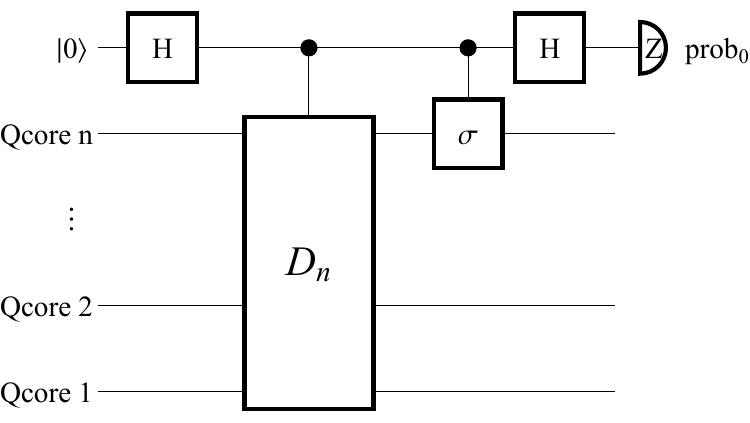}
		\caption{
            Circuit representation of the error mitigation technique: the quantum cores (Qcores) perform the same quantum computation independently in parallel. A derangement operation $D_n$ is then applyied immediately prior to measurement that uses the copies to validate each other. Errors in estimating the observable $\sigma$ are suppressed exponentially when increasing the number $n$ of cores. As each core consits of $N$ computational qubits, we show that the derangement circuit can be implemented efficiently with distributing $N (n-1)$ Bell-pairs between the quantum cores and performing $N (n-1)$ controlled-SWAP operations locally.
            This figure has been adapted from Ref.~\cite{koczorExponentialErrorSuppression2021}.
			\label{fig:derangement}
		}
	\end{centering}
\end{figure}

In particular, we use $n$ quantum cores to perform the same quantum computation in parallel. We use these (near) identical copies of the computational state to effectively verify each other via the derangement circuit as illustrated in \cref{fig:derangement}.
Following Ref.~\cite{koczorExponentialErrorSuppression2021}, if we entangle these copies with the derangement circuit and estimate the probability $\mathrm{Prob}_0$ that the ancilla qubit collapses into state $|0\rangle$, we can formally obtain the expectation value $\mathrm{Tr}[\rho^n \sigma]/\mathrm{Tr}[\rho^n]$.
This allows us to suppress errors in estimating the expectation value of an observable $\sigma$ exponentially in $n$. The main limitation of the approach is that a small coherent mismatch in the dominant eigenvector of the state $\rho$ may ultimately bias our estimates, however, this mismatch is exponentially less severe than the incoherent decay of the fidelity  \cite{koczorDominantEigenvectorNoisy2021a}.

Furthermore, the derangement circuits used to entangle the two registers are considerably shallower than typical computational quantum circuits used even in the context of near-term quantum algorithms \cite{koczorExponentialErrorSuppression2021}.
As such, the following three properties of the approach make it particularly relevant for on-chip multicore architecture designs. First, the two (or more) input quantum states $\rho_1$  and $\rho_2$ can be prepared completely independently in two physically separate cores.
Second, they are entangled with weak quantum links between the cores immediately prior to the ancilla measurement. Third, the entangling derangement operation is shallow: it decomposes into $N$ elementary controlled-SWAP operations between pairs of qubits, where $N$ is the number of computational qubits in the individual cores (registers).

We note that Ref.~\cite{koczorExponentialErrorSuppression2021} provides decompositions of derangement circuits for arbitrary $n$ into local, elementary entangling gates
whereas Ref.~\cite{hugginsVirtualDistillationQuantum2021} focused on the scenario of $n=2$ copies without requiring an ancilla qubit. In the following we outline an alternative implementation that is compatible with both the above techniques and utilises macroscopically separate quantum cores (which could be on a single chip for both methods, but could also be on multiple chips for the cavity-based method) with weak entangling links between them.

\begin{figure*}[th]
	\begin{centering}
		\includegraphics[width=\textwidth]{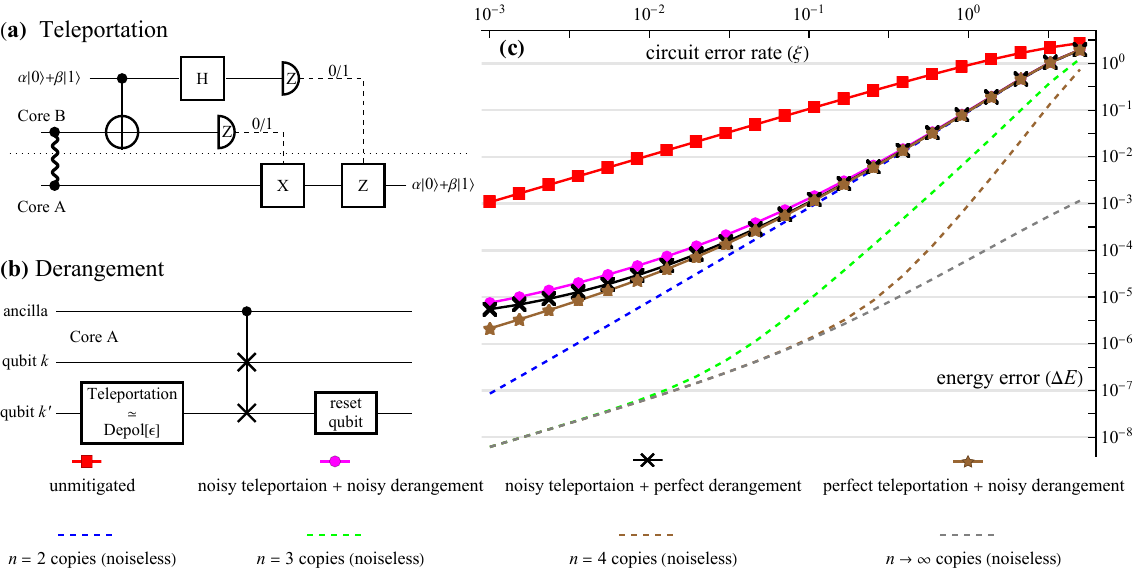}
		\caption{
			(a) Circuit for teleporting a single qubit state
			from core B to core A by consuming a Bell pair (wavy line) that had been distributed
			between the two cores
			(b) after teleporting the single-qubit we can locally
			apply the derangement circuit in core A, i.e., the controlled-swap operation between the
			two single qubits. The lower qubit is reset
			and repeating this procedure for all computational qubits $k$ implements the entire derangement circuit.
			Imperfections in the Bell pairs have a formally equivalent effect to a depolarising channel applied during
			the computation and therefore the derangement circuit mitigates these imperfections.
			(c) Numerical simulation of a spin-ring Hamiltonian. Error in estimating the ground state energy as a function of the expected number of errors in the state-preparation circuit ($\xi$). The unmitigated errors (red squares) are impressively removed by the error suppression technique (assuming noiseless derangement circuits) using $n$ copies of the computational state (dashed lines). In the practically relevant regime when $0.1 \leq \xi \leq 5$, the performance of the ESD/VD technique is similar to the ideal one even when taking into account imperfections in the derangement circuit (brown stars) or the imperfections in the Bell pair distribution (black crosses) or both (magenta circles).
			\label{fig:teleportation}
		}
	\end{centering}
\end{figure*}

Let us first note that distributing Bell pairs in principle enables universal quantum computation as it allows us to implement arbitrary long-range quantum-gates (via quantum gate teleportation). However, we choose an other way and we propose an approach that merely uses $N$ distributed Bell pairs to teleport single-qubit states between the two quantum cores. Let us explain now the technique on the specific example of $n=2$ copies. Refer also to \cref{fig:ill} for a schematic illustration of the process.

\emph{(1) Teleport single qubit state}---we aim to implement the quantum teleportation protocol in \cref{fig:teleportation}(a) to formally swap the state of a \emph{single} computational qubit from core 2 into a buffer qubit in core 1. This requires the preparation and distribution of a single Bell pair between the two cores as well as local operations and classical communication (LOCC transformations).

\emph{(2) Apply controlled-SWAP}---Once the single qubit state has been teleported into core 1, we can implement the controlled SWAP operation (the elementary building block of the derangement circuit) in core 1 locally as illustrated in \cref{fig:teleportation}(b).

The process of quantum-state teleportation (1) and consequent application of the controlled-SWAP operation (2) is then repeated $N$ times for all computational qubits. Note that during this process the buffer qubit in core 1 is always reset, while the computational qubits in core 2 are all measured out. Hence one of the copies of the computational state is destroyed, but this does not affect the measurement outcome of the ancilla qubit.

Our technique can be naturally generalised to the case of $n$ copies via a number of possible ways. For example, we may distribute Bell pairs between cores $(1-2), (1-3) \cdots (1-n)$ and use them to teleport $n$ single computational qubit states to buffer qubits in core $1$. We can then implement the controlled-derangement operator locally on $n$ copies of the single-qubit state and repeat the procedure $N$ times for all computational qubits.

The above approach has one significant advantage in the context of error mitigation:
We prove in \cref{sec:proof_pauli_error} that imperfections in the Bell-state preparation or measurement errors are guaranteed to result in a formal application of a single-qubit depolarising error channel to the individual computational qubits, as illustrated in \cref{fig:teleportation}(b). This is highly advantageous since imperfection in the long-range quantum teleportations are formally part of the state-preparation process and the derangement circuit is guaranteed to exponentially suppress these incoherent error contributions. 

Practical questions that arise in our setting are the following. (a) is the Bell-pair generation rate sufficient such that distributing $N$ Bell-pairs immediately prior to measurement is sufficiently fast when compared to the main computation? We can answer this affirmatively given the distribution of a Bell pair
is comparable to the time of local operations -- which we expect is the case for both the cavity-based and shuttling alternatives.
(b) can `memory qubits' be manufactured that buffer all $N$ Bell-pairs thereby enabling them to be generated in parallel with the main computation?

We numerically simulate the present protocol in the following using realistic error rates of Bell pairs as we have established above.

\subsection{Numerical simulations}

We simulate a variational quantum eigensolver (VQE) application and consider a spin-ring Hamiltonian with a constant coupling $J=0.1$ and uniformly randomly generated on-site interaction strengths $\omega_k \in [-1,1]$ as 
\begin{equation}\label{spin_ring}
	\mathcal{H} =  \sum_{k \in \text{ring}(N)} \omega_k Z_k  + J \, \vec{\sigma}_k \cdot \vec{\sigma}_{k+1},
\end{equation}
as relevant in the context of manybody localisation \cite{nandkishoreManyBodyLocalizationThermalization2015, luitzManybodyLocalizationEdge2015, childsFirstQuantumSimulation2018}.

We explicitly simulate $N=6$ qubits and $n=2$ copies (equivalent to a 26-qubit pure-state simulation) and we assume the ground state is prepared via a variational Hamiltonian ansatz of $l=20$ layers to a precision of $10^{-4}$ \cite{farhiQuantumApproximateOptimization2014,babbushLowDepthQuantumSimulation2018, weckerProgressPracticalQuantum2015,cadeStrategiesSolvingFermiHubbard2020, wiersemaExploringEntanglementOptimization2020}. Refer to \cref{app:num_sim} for more details.

We simulate quantum cores that can locally implement parametrised controlled-$Z$ entangling gates as well as single qubit rotation gates and we assume the error rate of single-qubit gates are 5 times smaller than that of the entangling gates. While such a ratio is very common in experimental systems, we do not intend to capture exact noise characteristics of state-of-the-art entangling gates and simply note that the literature is evolving rapidly \cite{xueQuantumLogicSpin2022}.
Independently of the gate error rates, the main computation requires overall $3Nl = 360$ applications of local entangling gates. Furthermore, we adapt techniques of \cite{koczorExponentialErrorSuppression2021} for implementing the derangement circuit using local entangling gates:  implementing the derangement circuit requires $5 N = 30$ applications of local entangling gates as well as the distribution of $N=6$ Bell pairs. We can therefore expect that due to its modest resource requirements the derangement circuit is much less affected by gate noise than the main computation will be.

We simulate the approach assuming  noisy Bell pairs of fidelity $f$ that have been prepared via the long range entangling links outlined above.
In the following we assume that $f = 99.5\%$ which we have shown above is reachable with current state-of-the-art technologies but we can also expect this figure will improve with future hardware developments.
As discussed above, we efficiently model the process of teleporting single qubit states by formally applying single-qubit depolarising noise of probability $0.5\%$ after every qubit in one of the copies of the computational quantum states.

\cref{fig:teleportation} shows energy estimation errors as a function of the number of expected errors (circuit error rate $\xi$) in the state-preparation circuit. Since the measurement cost of error mitigation techniques generally increase exponentially with the circuit error rate, we focus on the practically most important regime as $0.1 \leq \xi \leq 5$. In this regime, the unmitigated errors (red squares) are significantly reduced even when we take into account the imperfections in the derangement circuit (brown stars) and imperfections in the Bell-state preparation (black crosses) or both (magenta circles).
The error due to imperfect Bell pairs (black crosses) approaches a very small constant error for $\xi \rightarrow 0$. This error is due to a small coherent shift in the dominant eigenvector of the quantum state $\rho$ introduced by the formal application of $N$ single-qubit depolarising noise channels immediately after the main computation.

While these simulations confirm that the ESD/VD technique is impressively robust to imperfections in the long-range entangling links, they do come at an increased measurement cost.  In particular, since imperfections in the $N$ distribted Bell pairs are equivalent to local depolarising channels, we estimate that the probability that no error happens during the long range teleportation is $f^N$, where $f$ is the fidelity of the Bell pairs. While the ESD/VD technique filters out erroneous contributions, the probability of an error-free outcome is attenuated and therefore the expectation value requires an increased number of samples to be resolved to a sufficient accuracy. For example, in the present case of $N=6$ qubits we find the probability of no errors occuring during teleportation is $0.995^6 \approx 0.97$ not significantly attenuated. However, when scaling up computations this probability decreases exponentially with $N$. Nevertheless, even with, e.g., 100 qubits, we can still estimate the encouraging probability $0.995^{100} \approx 0.60$. 

In summary, our numerical simulations confirm that the ESD/VD technique is indeed compatible with the weakly connected multicore concept and imperfections in the long-range links are impressively well mitigated by the derangement circuits.

\subsection{Other Applications}
Besides exponential error suppression, there are a number of other problems that can be efficiently implemented using our modular architecture. 

The simplest example is the SWAP-test whereby we prepare two different states $\ket{\psi_1}$ and $\ket{\psi_2}$ in two input registers (quantum cores) and measure their overlap $|\langle \psi_1 | \psi_2 \rangle|^2$ using the controlled-SWAP operation. This operation is directly analogous to what we have done for the 2-copy error mitigation scheme in \cref{fig:derangement}. The SWAP-test is an elementary subroutine crucial for the implementation of a number of
important algorithms, which include the following.
(a) finding excited states of quantum systems, such as in quantum chemistry \cite{higgottVariationalQuantumComputation2019,jonesVariationalQuantumAlgorithms2019};
(b) Simulating quantum dynamics of mixed quantum states and general processes \cite{yuanTheoryVariationalQuantum2019,endoVariationalQuantumSimulation2020};
(c) implementing the quantum natural gradient optimisation approach in variational 
quantum eigensolvers and in other variational quantum algorithms \cite{koczorQuantumNaturalGradient2020,koczorQuantumAnalyticDescent2022,mariEstimatingGradientHigherorder2021}. 

Our modular architecture is also well-suited for performing simulations for problems with a Hamiltonian that is modular in nature (has clusters of subsystems). In these cases, the ansatz for the variational algorithm of such problems can be implemented natively on our architecture for efficient simulations. There are many interesting physical problems of this kind~\cite{yuanQuantumSimulationHybrid2021}, including problems in chemistry~\cite{garlattiPortrayingEntanglementMolecular2017,timcoEngineeringCouplingMolecular2009}, many-body physics~\cite{kitaevUnpairedMajoranaFermions2001,sauNonAbelianQuantumOrder2010}, quantum field theory~\cite{jordanQuantumAlgorithmsQuantum2012,jordanQuantumComputationScattering2014}, and quantum gravity~\cite{maldacenaDivingTraversableWormholes2017,maldacenaEternalTraversableWormhole2018}. Supporting the prospects for successful quantum advantage in such tasks, there are studies anticipating the challenges of compiling onto target hardware with specific topologies~\cite{akibueNetworkCodingDistributed2016,andres-martinezAutomatedDistributionQuantum2019,ferrariCompilerDesignDistributed2021} and examinations of whether modules of modest size can `punch above their weight' in simulating more complex quantum systems~\cite{bravyiTradingClassicalQuantum2016,pengSimulatingLargeQuantum2020,tangCutQCUsingSmall2021}.
However, the requirements towards the quality of the communication channels may depend entirely on the structure of the problem.

It is also possible to implement quantum error correction codes on such a modular architecture. The simplest example would be code concatenation, in which each module will implement a base code, and the long range links among the modules will implement another code at the logical level of the base code. A more concrete example would be the modular way to implement large scale surface code studied in Ref.~\cite{liHierarchicalSurfaceCode2016}. They found a high error threshold for the noise in the long-range connections among the modules even without purification. In the limit of small modules of a few qubits or a few tens of qubits, there are detailed studies of the most efficient means of supporting fault tolerant codes~\cite{nickersonFreelyScalableQuantum2014,nigmatullinMinimallyComplexIon2016}. More recently, Ref.~\cite{tremblayConstantOverheadQuantumError2022} established that it is possible to perform constant overhead quantum error correction in a 2D architecture with non-crossing long range connections. Though this is not a modular architecture, the long range gates that we have studied in this article will still be highly relevant to their practical implementation in silicon.

\section{Discussion and Conclusion}

In this work we have investigated the practicality and utility of an interlinked multicore architecture for quantum computing. We have focused on silicon-based devices where the paradigm is particularly relevant: it is natural to consider a large number of independent, noisy quantum cores due to their inexpensive fabrication process and the small surface area required by a given modest-sized processor.
Recognising that inter-core operations may be fundamentally more noisy than intra-core processes, we eschewed the goal of performing direct unitary two-qubit quantum gates between remote qubits. Instead we focused on distributing Bell pairs between the processors. This allowed us to use `weak', noisy entangling links for the purpose of generating raw Bell pairs, which can then be upgraded to far higher fidelities using powerful entanglement distillation techniques. 

We briefly reviewed the experimental literature on physical realisations of potential `quantum links' that have already been demonstrated in state-of-the-art experiments: We identified cavity- and shuttling-based techniques as the most promising ones for our proposal. We have comprehensively explored advantages and limitations of both techniques in terms of numerical simulations. In our modelling we identified and accounted for the leading error and noise contributions in the long-range entangling links. As such, even when assuming parameter regimes that are realistic in current experiments, our numerical simulations confirm that both techniques (in combination with entanglement distillation) would support Bell pair generations between distant quantum cores with fidelities about $99.5 \%$.
Our simulations also confirmed that slightly increasing the number of rounds in entanglement distillation allows us to obtain near-perfect Bell pairs approaching error levels of local, intra-core operations.
To inform our models with realistic parameters, such as relaxation rates or gate and measurement times, we have used the best reported experimental values; even though there is no single experiment that would achieve all those values at once, we note that they are very much moving targets and the numbers are constantly improving.

We reviewed a number of potential applications which build on a modular computational architecture. Most importantly, we identify that the recently introduced Error Suppression by Derangements \cite{koczorExponentialErrorSuppression2021} and Virtual Distillation \cite{hugginsVirtualDistillationQuantum2021} techniques fit well with our proposal, since these error mitigation schemes assume that $n$ copies of a quantum computation are performed independently in separate cores. Through our long-range links, the copies are entangled via a derangement operation which bridges the $n$ computational cores and this allows us to mitigate errors exponentially in $n$. The great advantage of our proposal is that the long-range links are formally part of the main computation and thereby all experimental imperfections are mitigated by the derangement operation.

We numerically simulated such a practically relevant application: a VQE solving the ground state of a spin system. These simulations confirm that even with Bell pairs of fidelity $99.5 \%$, as achievable with current technology, we can impressively suppress errors of local quantum gates in the main computation and thereby obtain accurate expectation values of observables in practical settings. We have also identified a number of other potential applications, such as the SWAP test, which are highly relevant in the context of both near-term and error-corrected quantum computations.

We finally conclude that long-range entangling operations that have already been realised in experiments can be used for linking a multitude of silicon-based computational quantum cores. Such a modular architecture would enable a variety of near-term applications and, in particular, would enable powerful error mitigation techniques to be implemented in silicon devices. 
While in the present work we considered silicon architectures, we remark that these concepts naturally generalise to other platforms as well: distributing and purifying Bell pairs between quantum hardware will enable powerful error mitigation and other applications in other platforms via the techniques outlined in this work.

\emph{Note added}---We note that recently a paper appeared [arXiv:2202.11793] that
investigates in great detail shuttling arrays as in the present work and reaches
even more encouraging fidelity estimates. During the process of finalising this publication
a paper appeared [arXiv:2208.11151] that investigates cavity-based links between silicon spin qubits in the resonant regime similarly as in the
present work and finds similar conclusions.

\section*{Acknowledgements}
The authors thank John Morton, Fernando Gonzalez-Zalba, Sofia Patomäki, and Felix von Horstig for useful discussions. 
The authors acknowledge support from Innovate UK project 133997: Multicore NISQ Processors on Silicon Chips, from the EPSRC QCS Hub EP/T001062/1, from EU H2020-FETFLAG-03-2018 under the grant agreement No 820495 (AQTION) and from the IARPA funded LogiQ project. 
B.K. acknowledges financial support from the Glasstone Research Fellowship of the University of Oxford. 
Z.C. is supported by the Junior Research Fellowship from St John’s College, Oxford. This research was supported by European Union’s Horizon 2020 research and innovation programme under Grant Agreement No. 951852 (QLSI).

\appendix

\section{Cavity}
\subsection{\label{app:parameter_regime} Parameter regime}

In our modelling we have explicitly---and numerically exactly---simulated the unitary dynamics generated by the Hamiltonain $H$ in \cref{eq:two_qubit_hamiltonian} in a resonant parameter regime, i.e., we did not use a rotating frame approximation.
To ensure numerical stability we numerically exactly exponentiated the matrix $-i t H$ after projecting the
bosonic creation and annihilitation operators into the subspace spanned by the first seven Fock states (eigenstates with a fixed
number of photons). Furthermore, we computed the matrix $-i t H$ explicitly assuming a set of dimensionless parameters
and the resulting optimised values are reported in \cref{tab:cavity_parameters}.

Recall that given a Hamiltonian $H = a H_1 + b H_2$ in terms of the dimensionless parameters $a$ and $b$ we can map the dynamics under the dimensionless time $t$ to any physical set of parameters via the symmetry
\begin{equation*}
    e^{-it (a H_1 + b H_2)}
    =
    e^{-i t/\lambda (\lambda a H_1 + \lambda b H_2)}
    =
    e^{-i t' (a' H_1 + b' H_2)}.
\end{equation*}
Here we have introduced a new set of parameters
$a'=\lambda a$, $b'=\lambda b$ and $t'=t/\lambda$, and the \emph{arbitrary} scaling factor $\lambda$. We choose a possible scaling factor $\lambda:=\SI{100}{\mega\hertz}$ such that the resulting optimal phyisical parameters fall into a regime that can be realised with near-term technology (see \cite{ibbersonLargeDispersiveInteraction2021} for the possibility of reaching large charge-cavity couplings). Note, however, that the optimal unitary dynamics can be mapped to any other regime, e.g., choosing $\lambda:=\SI{200}{\mega\hertz}$ allows a faster generation of Bell pairs ($t' = \SI{7.5}{\nano\second}$) but requires the fabrication of twice as large couplings (and all other dymanical parameters twice as large).
\begin{figure}[tb]
	\begin{centering}
		\includegraphics[width=0.45\textwidth]{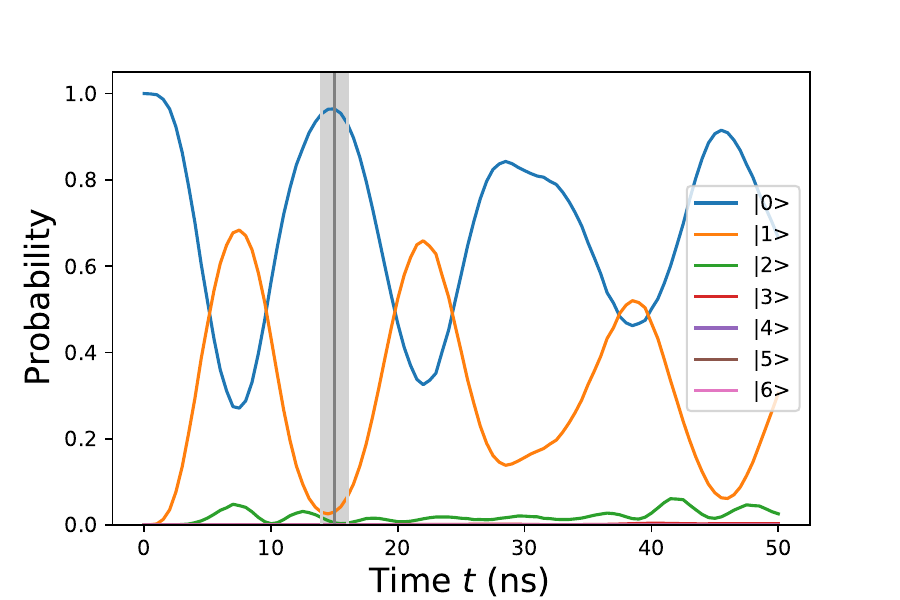}
		\caption{Evolution of the photon's state after a successful charge measurement. When we stop the interaction and make the measurements, there is a $\approx 0.96$ probability to find the cavity in the vacuum state.}
		\label{fig:photon_proba}
	\end{centering}
\end{figure}
\begin{table}[tb]
    \begin{ruledtabular}
    \begin{tabular}{lccc}
    description & symbol & unitless & physical value  \\[1mm]
    \hline
    \\[-3mm]
        magnetic field & $B$ & $100$ & $10^4$\,MHz\\[1mm]
        transverse magnetic field& $b_{x,1}$ & $5.26$ & $526$\,MHz\\[1mm]
        transverse magnetic field& $b_{x,2}$ &$1.74$ & $174$\,MHz \\[1mm]
        charge cavity coupling & $g_{c,1}$  &$3.9$ & $390$\,MHz \\[1mm]
        charge cavity coupling &$g_{c,2}$ & $1.4$& $140$\,MHz \\[1mm]
        DQD detuning & $\epsilon_1$ & $0$& $0$\,MHz \\[1mm]
        DQD detuning & $\epsilon_2$ &  $0$& $0$\,MHz  \\[1mm]
        optimal time & $t$ &  $1.5$ & $15$\,ns  \\[1mm]
    \end{tabular}
    \end{ruledtabular}
    \caption{\label{tab:cavity_parameters}
	Parameters used in the cavity model. These parameters were obtained from a gradient-based optimisation. $B$ is the parallel magnetic field used to lift the spin degeneracy, $b_{x,1}$ ($b_{x,2}$) is the transverse magnetic field above the first (second) DQD which hybridises the spin and charge degrees of freedom, $g_{c,1}$ ($g_{c,2}$) is the dipole coupling constant between the cavity and the first (second) DQD and $\epsilon_1$ ($\epsilon_2$) the detuning.} 
\end{table}

While the arbitrary choice of $\lambda$ is an exact symmetry of the unitary dynamics, we can introduce a pseudosymmetry. As discussed above, we explicitly take into account and simulate the effect of the magnetic field $B$ which is by orders of magnitude larger than all other dynamical parameters. Given that one could make the rotating frame approximation and remove this interaction without significantly affecting the dynamics, in numerical simulations we have confirmed that indeed increasing or decreasing the value of $B$ by a small factor, e.g., $2$ or $3$, does only affect the resulting dynamics to a very small extent. This ensures us that an experimental system is robust to the choice of the $B$ as long as the resonant condition is satisfied.

	\begin{figure*}
		\centering
		\includegraphics[width=0.48\textwidth]{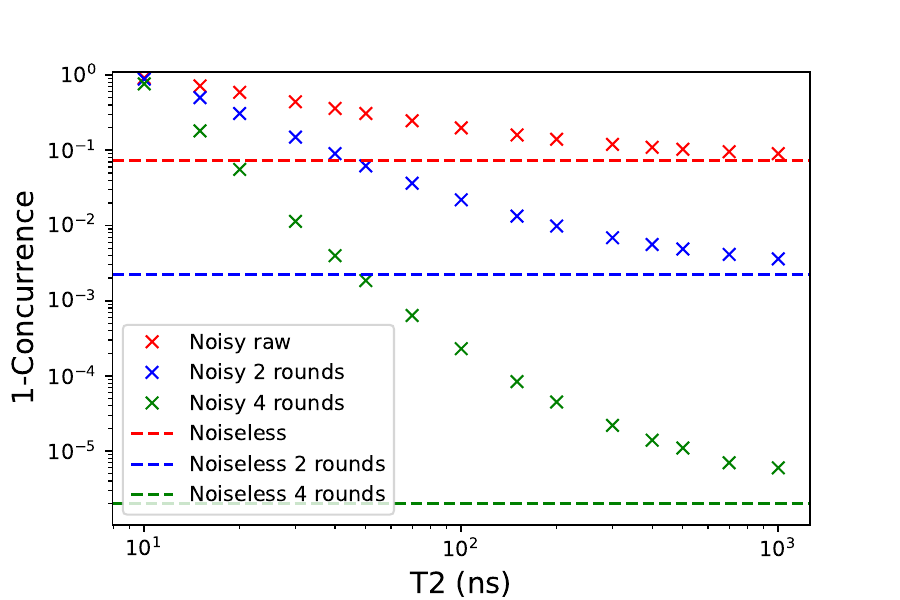}
		\includegraphics[width=0.48\textwidth]{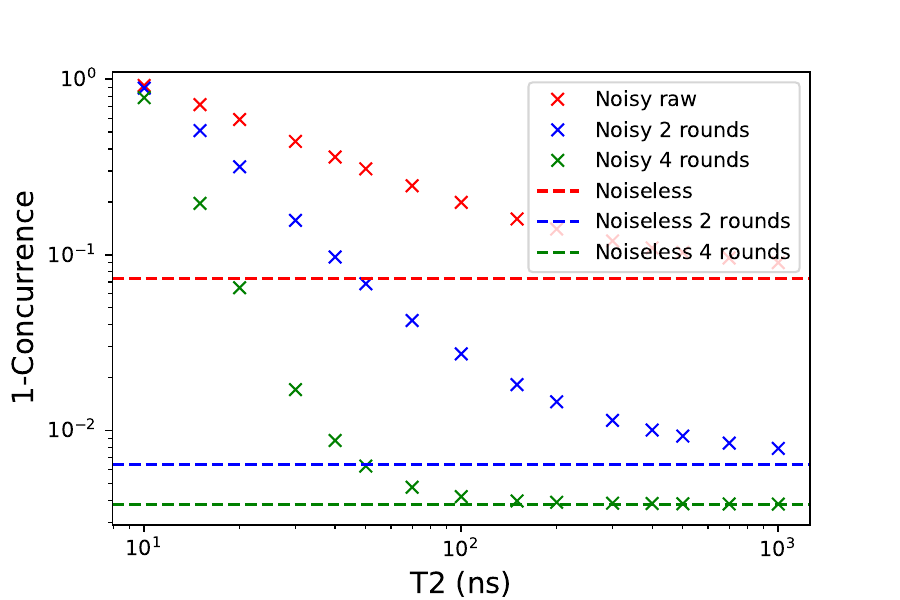}
		\caption{Concurrence (measure of entanglement) as a function of charge decoherence for different levels of purification. Left (right) without (with) errors in the local quantum gates in the purification process.
			The dashed red line represents a perfect unitary evolution (no decoherence) and perfect charge measurements  and the only source of imperfections is the fact that our fast, resonant dynamics in \cref{eq:two_qubit_hamiltonian} cannot exactly generate a Bell pair.}
		\label{fig:purification_one_phase_three_s}
	\end{figure*}

\subsection{\label{sec:photon_after_meas} State of the cavity after charge measurements}

To be able to reuse the cavity straight after having generated a Bell pair, we need to end up the protocol with the photon in the vacuum state. One possibility would be to measure the charges and the cavity, and deem the result as successful if we are both measuring the charges in either the state $\ket{\text{LR}}$ or $\ket{\text{RL}}$ and the cavity in the vacuum state. However, this would have led to a decrease of the success probability compared to the case when we only perform charge measurements. Fortunately, after a successful charge measurement, the probability to find the resonator in the vacuum state is high $\approx 0.96$ (see \cref{fig:photon_proba}). Even though the probability to be in the zero state for the cavity is high, one may need to consider the cavity heating up during consecutively repeated Bell pair generations. Nevertheless, given cavities in current experiments have typical decay rates $\approx 1 \text{MHz}$ \cite{borjansResonantMicrowavemediatedInteractions2020}, dissipation should occur on timescales that is smaller than a single round of entanglement purification.
Furthermore, the cavity is only used for a very small fraction of the procedure -- allowing for cooling protocols to 
be periodically performed if necessary.

\subsection{Lindblad master equation}
The evolution of an $d$-dimensional quantum system in the state $\rho$ with time is given by the Lindblad master equation,
\begin{align}
\label{eq:lindblad_master_equation}
    \frac{\partial \rho(t)}{\partial t} &= \mathcal{L}\rho   \\\nonumber
    &=  -i\left[H,\rho\right] + \sum_{i}\gamma_{i}\left(A_i\rho A_i^{\dagger}-\frac{1}{2}\left\{ A_{i}^{\dagger}A_{i},\rho\right\}\right),
\end{align}
where $\gamma_i$ are positive rates, and $A_{i}$ Lindblad terms. Given the Hamiltonian is time independent we can exponentiate the superoperator $\mathcal{L}$, and the above equation can be rewritten as
\begin{align}
    \rho(t) = e^{\mathcal{L}t}\rho(0).
    \label{eq:lindblad_time_evol}
\end{align}
We can use standard numerical techniques for the exponentiation \cite{am-shallemThreeApproachesRepresenting2015} if we represent $\rho$ and $\mathcal{L}$ as a vector and a matrix respectively. To do so, we need to flatten $\rho$, meaning that we will stack the different rows one after the other. $\rho$ becomes then a $1\times d^2$ vector. The superoperator $\mathcal{L}$ becomes a $d^2\times d^2$ matrix, by applying the following rules,
\begin{align}
A\rho &\rightarrow \mathbb{I}\otimes A\rho \nonumber \\
\rho B &\rightarrow B^{T}\otimes\mathbb{I}\rho\nonumber \\
A\rho B &\rightarrow B^{T}\otimes A\rho,
\end{align}
where on the left-hand side we have the operation in the operator formalism and on the right-hand side the equivalent operation in the matrix-vector formalism.

\cref{eq:lindblad_master_equation} becomes then,
\begin{align}
    &\frac{\partial \rho(t)}{\partial t} = -i\left( \mathbb{I}\otimes H- H^{T}\otimes\mathbb{I}\right)\rho + \sum_{i}\gamma_{i} \times  \\
    &\left( \left(A_{i}^{\dagger}\right)^{T}\otimes A_{i}-\frac{1}{2}\mathbb{I}\otimes A_{i}^{\dagger}A_{i}-\frac{1}{2}\left(A_{i}^{\dagger}A_{i}\right)^{T}\otimes\mathbb{I}\right)\rho \nonumber.
\end{align}

\subsection{Time evolution}
There are two ways to numerically simulate \cref{eq:lindblad_time_evol}. One way consists in recomputing the exponential $e^{\mathcal{L}t}$ for each time $t$ and apply it to $\rho(0)$. We discarded this method as exponentiating $\mathcal{L}$, which is represented by a $(12544\times12544)$ matrix, is time-consuming. The other way relies on time discretization. In this case, as we only compute once the exponential for a small time step $dt$, and apply it sequentially to propagate the state, the simulation is much faster. 
For large time $t$, the first method is more accurate. However, in our case both techniques give similar results, reinforcing our choice to use the second method.

\subsection{State prior to measurement} \label{app:state_analysis}

We start the dynamics with the initial quantum state as $| \psi_{init} \rangle = |{--}\rangle_{ch} |{\uparrow\uparrow}\rangle_{sp} |0\rangle_{cav}$,
such that the charge degrees of freedom are in the separable state $|{--}\rangle_{ch}$,
the spin degrees of freedom are in the separable state  $|{\uparrow \uparrow}\rangle_{sp}$
and the cavity is separable in the vacuum state $|0\rangle_{cav}$.

\begin{figure}[tb]
	\includegraphics[width=0.47\textwidth]{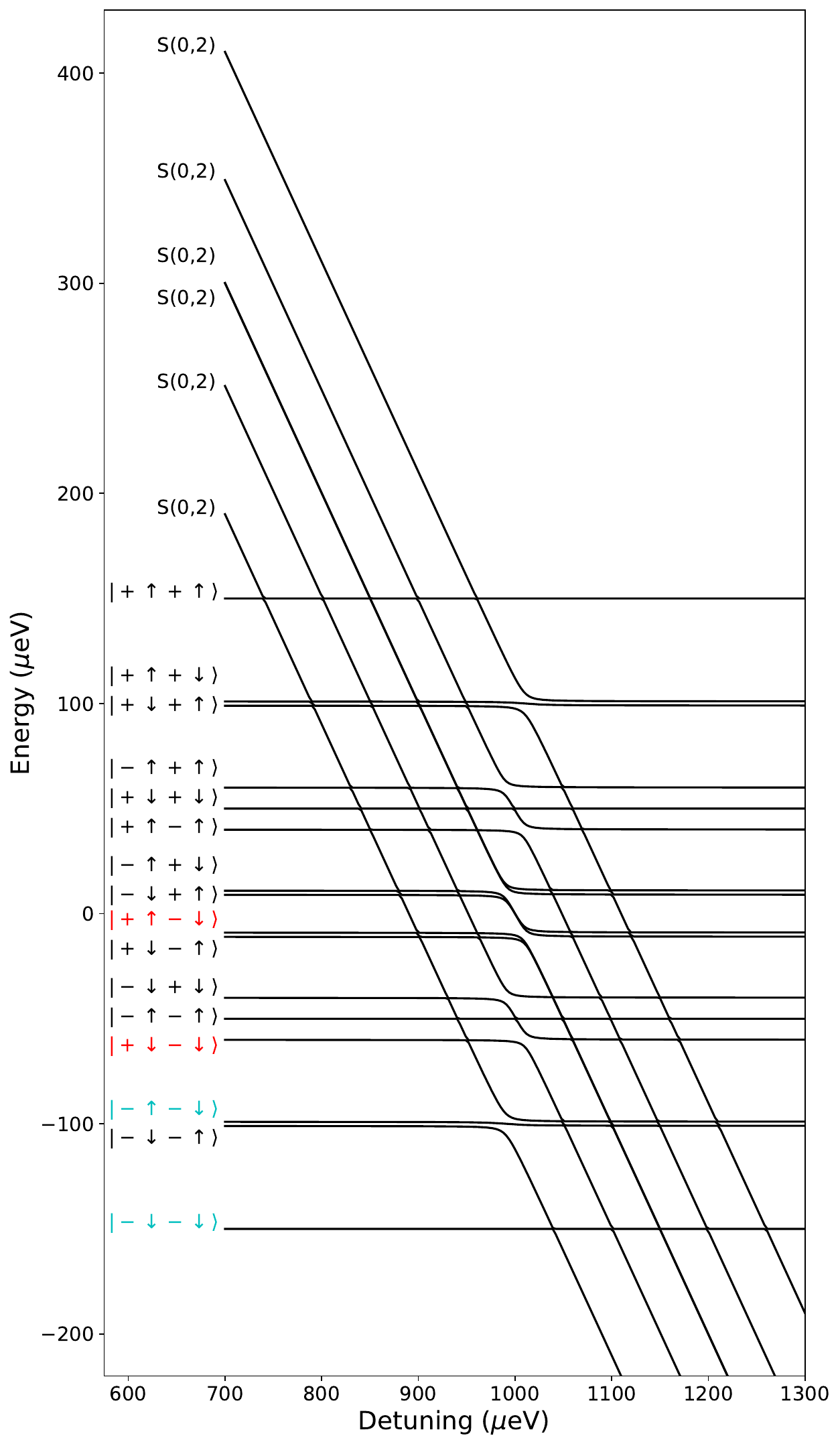}
	\caption{
		See caption on the right.
	}
	\label{fig:valley_measure}
\end{figure}

\addtocounter{figure}{-1}
\begin{figure}[tb]
	\caption{
		\label{fig:energy_levels}
		Energy-level diagram of $H_{2e}$ from \cref{eq:H_2eDQD} in the positive-detuning region where electrons only occupy the ground orbital wavefunction. Here, $U = 1$~\SI{}{\milli\eV}, $t = 5$~\SI{}{\micro\eV}, $E_{Z,L} = 51$~\SI{}{\micro\eV}, $E_{Z,R} = 49$~\SI{}{\micro\eV}, $E_{V,L} = 90$~\SI{}{\micro\eV}, and $E_{Z,R} = 110$~\SI{}{\micro\eV}. In this experimentally realistic regime, $\Delta E_Z > 0 > \Delta E_V$ and $|\Delta E_V|>|\Delta E_Z|$, although the Zeeman difference here has been exaggerated to enhance the visibility of distinct traces. The state labels are ordered as $\ket{vs}_L\ket{vs}_R$. The highlighted labels indicate the 4 states where the right dot is initially populated with an ancilla ground-valley/ground-spin state, and the left dot is populated with a shuttled spin-valley state. Of these 4 states, only those with the left dot electron in the excited valley state (red) will adiabatically evolve to (0,2), while the states in the ground valley (blue) will remain in (1,1). Therefore, such a double-dot arrangement along with a proximal charge sensor can be used to perform a valley-to-charge projective measurement while preserving the spin state.
	}
\end{figure}

Given we assume a resonant condition
with the parameters $B_{1} = B_{2} = \omega_r \equiv B $ as well as $2t_{c,1} = 2t_{c,2} = 2t_c \equiv B$ with no detuning $\epsilon=0$,
the expected energy of this initial state is exactly $E=0$.
Furthermore, for any other state we obtain from this initial state by
pairs of  ``flips'', e.g., $|{-+}\rangle_{ch} |\uparrow \downarrow  \rangle |0\rangle_{cav}$,
we obtain an identical energy $E=0$ given any flip represents an energy quantum of $\pm B$.
Assuming no decoherence, at our optimal evolution time in \cref{fig:prob_concurrence}
we obtain the quantum state
\begin{equation}\label{eq:ideal_st}
	| \psi \rangle  = \sqrt{ F }  \Big(
	| \Psi^- \rangle_{ch}
	| \Psi^- \rangle_{sp}
	|0\rangle_{cav}
	\Big)
	 + \sqrt{1-F} | \psi_{err} \rangle,
\end{equation}
where with a high fidelity $F\approx 0.89$ we obtain our desired entangled state 
with a Bell pair $| \Psi^- \rangle = (|{\uparrow\downarrow}\rangle{-}|{\downarrow\uparrow}\rangle)/\sqrt{2}$
between  the two spins
up to some coherent error $| \psi_{err} \rangle$.

As we discussed in \cref{sec:bell_gen}
this coherent error is present because we have fewer parameters of control in the Hamiltonian
than the number of constraints the states need to satisfy in order be usefully entangled.
Given the desired state above has energy $E=0$, as it can formally be obtained by an even number of energy flips
from the initial state, the coherent error state $| \psi_{err} \rangle$ must also have energy $E=0$
due to energy conservation under a unitary evolution. While $| \psi_{err} \rangle$ is a complex quantum 
state that is a superposition of nearly all charge and spin configurations, we find that the dominant contribution
in $| \psi_{err} \rangle$ is the state
\begin{equation}\label{eq:dominant}
	| \Phi^- \rangle_{ch}
	|{\downarrow\downarrow}\rangle_{sp}
	|1\rangle_{cav},
\end{equation}
which occurs with a probability $F' \approx 0.05$ as the fidelity with respect to the full state $| \psi \rangle$. 
Indeed this state has energy $E=0$, however,
given the charge is entangled as $| \Phi^- \rangle_{ch} = (|LL\rangle{-}|RR\rangle)/\sqrt{2}$,
energy conservation forces the spins into the separable state $|{\downarrow\downarrow}\rangle_{sp}$
while the cavity also obtains a photon.

This motivates our measurement-based protocol: we measure the cavity state and only accept the outcomes
$|LR\rangle$ or $|RL\rangle$ which does not affect the ideal contribution of the state in \cref{eq:ideal_st},
however, it eliminates the dominant coherent error contribution from \cref{eq:dominant}. This necessarily increases entanglement
between the spins given the spins are separable in the error contribution in \cref{eq:dominant} that we eliminate. Furthermore, the
probability of an empty cavity is also increased given the state we project out has a non-empty cavity.

\begin{figure*}[tb]
	\centering
	\includegraphics[width=0.25\textwidth]{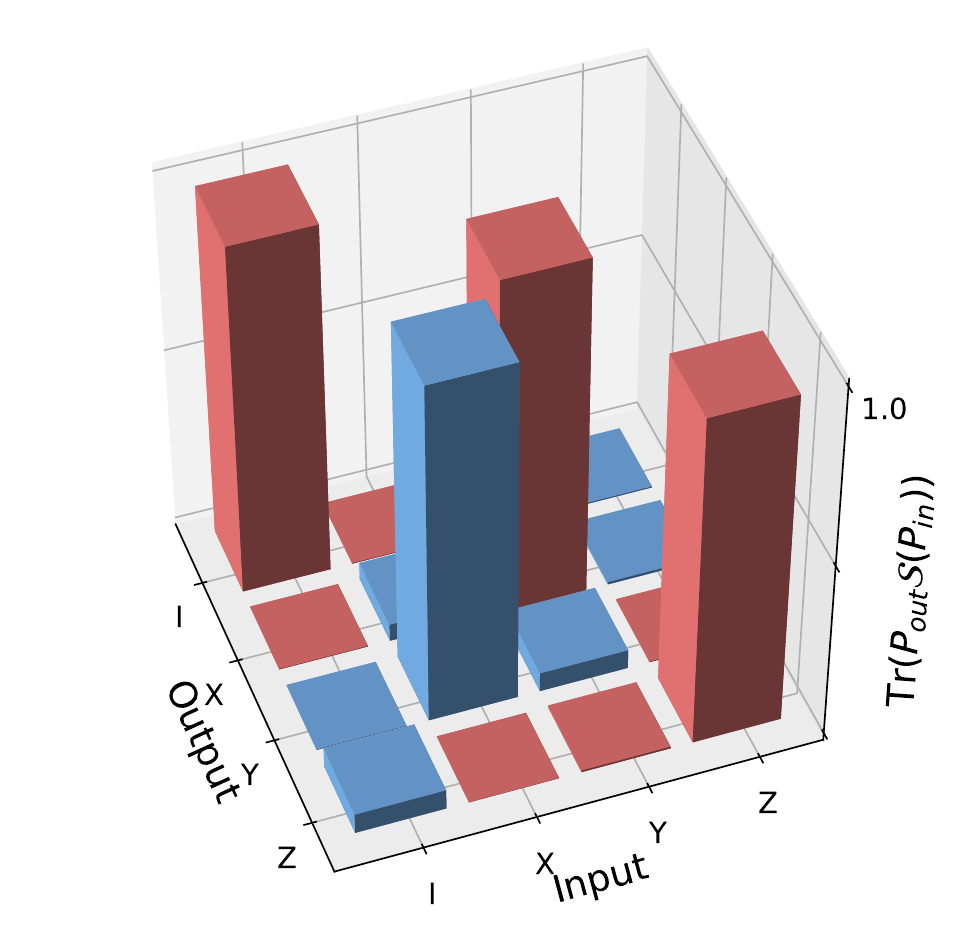}
	\includegraphics[width=0.35\textwidth]{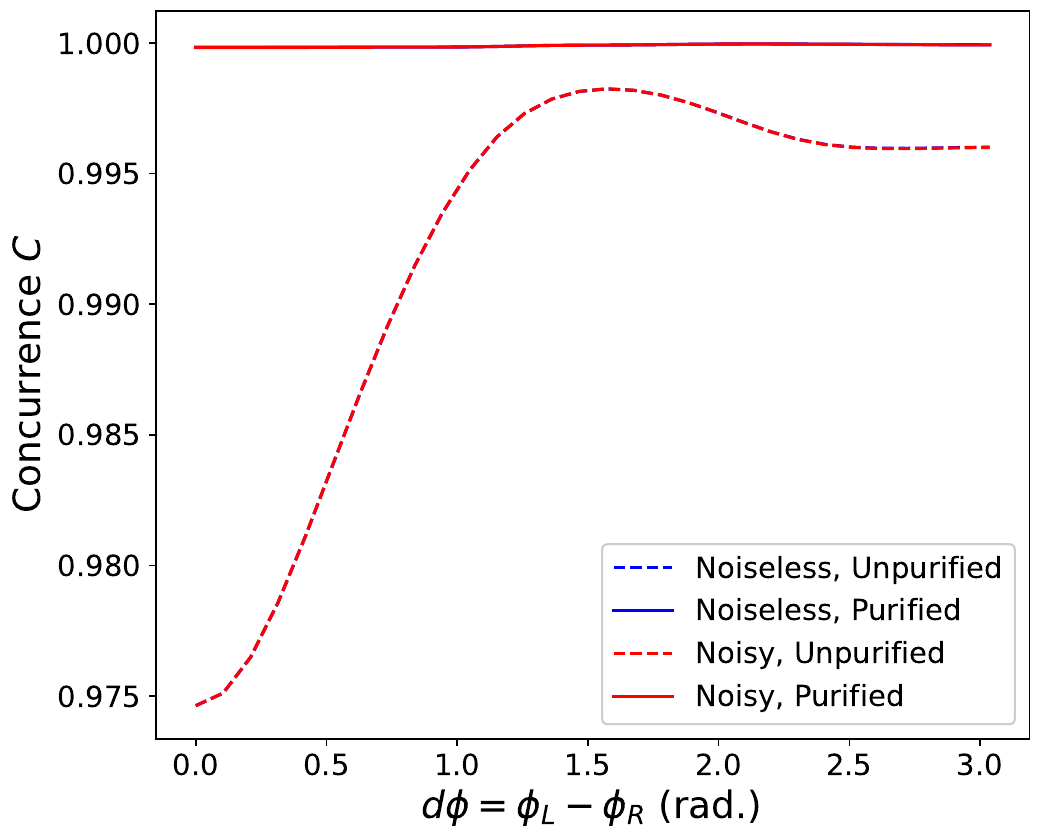}
	\includegraphics[width=0.35\textwidth]{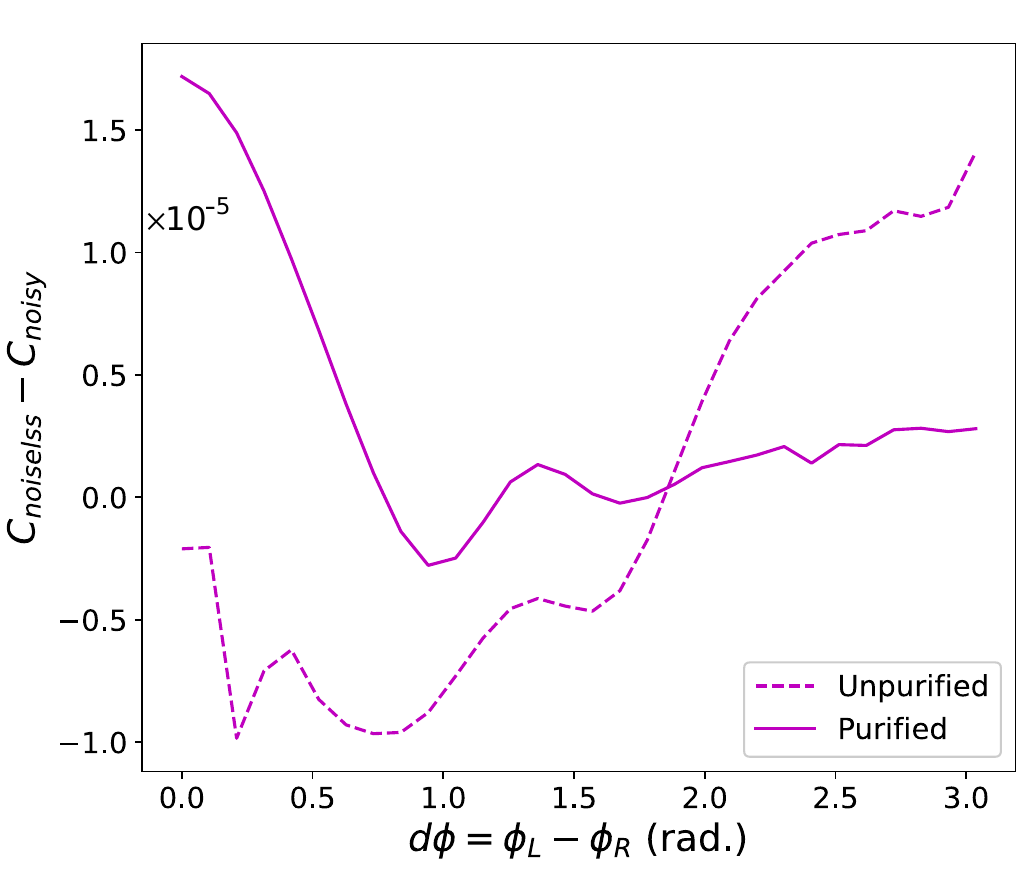}
	\caption{(left)
		Pauli transfer matrix acting on the shuttled spin, corresponding to the 25-dot shuttling channel $\mathcal{S}$ shown in (a) when $SD_\phi=\pi/4$ and $t_c=$~\SI{30}{\micro\eV}. Red bars indicate positive matrix elements, while blue bars indicate negative matrix elements.
		(middle) Calculation of the unpurified and purified spin-spin concurrences when shuttling through a DQD with a bare tunnel coupling of $t_c=$~\SI{20}{\micro\eV} and equivalent valley splittings of $E_V = 2|\Delta| =$~\SI{150}{\micro\eV}. $b_x=b_z=$~\SI{1}{\micro\eV} and $B=$~\SI{40}{\micro\eV}. The tunnel coupling has been decreased in order to maximize the effect of charge noise. (right) The difference between noisy and noiseless simulations shows charge noise has a negligible effect on the concurrence at the tunnel couplings and sweep speeds of interest.}
	\label{fig:shuttling_charge_noise}
\end{figure*}

\subsection{\label{app:alt_purification} Alternative purification protocol}
By adding single S gates to the three last rounds of purification, one can obtain a better purification scheme than the one presented in the main text for the case of the cavity-based approach. In \cref{fig:purification_one_phase_three_s}, we note that the concurrence is slightly improved 
in the ideal case. However, in the noisy case, we increase the noise (and thus lower the concurrence) by adding these new gates. With better hardware, one would transition from the scheme presented in the main text to this one.

\subsection{\label{app:generation_rate} Average generation rate}

Assuming a raw Bell pair is generated with time $T_{raw} = 15 ~\SI{}{\nano \second}$ and we neglect gate and measurement times in the
purification process, the time to generate a
purified Bell pair is $T_{raw} N_{avg}$ where the average number of required raw Bell pairs is $N_{avg}$. 
More realistically, the average time for generating a single purified Bell pair is
\begin{align*}
T_{avg} = T_{raw} \, N_{avg} &+ T_{1 qb}  \,  \frac{ N_{1 qb, avg} }{4}\\
&+T_{2 qb} \,  \frac{N_{2 qb, avg} }{ 2 } + T_{meas}  \,  \frac{ N_{meas, avg}  } { 2 },
\end{align*}
which depends on the number of single-qubit gates $N_{1qb, avg} = 13.6$, on the number of two-qubit gates and on measurements $N_{2qb,avg} = N_{meas, avg} = 6.8$.
In these figures we assumed $T_{2, c} = 400 ~\SI{}{\nano \second}$
and we have divided $N_{1 qubit, avg}$ by $4$ and $N_{2 qubit, avg}, N_{meas, avg}$ by 2 as the four single-qubit gates, the two two-qubit gates and the two measurements used in a purification round can be performed in parallel.
The generation rate is then given by $g_{Bell} = [T_{avg}]^{-1}$.
Upon substituting typical state-of-the-art values for
one-, two-qubit gate and measurement times as $T_{1 qubit} = 1 ~\mu s, \; T_{2 qubit} = 100 ~ \SI{}{\nano \second}, \; T_{meas} = 1 ~\mu s$
\cite{xueCMOSbasedCryogenicControl2021,noiriFastUniversalQuantum2022,connorsRapidHighFidelitySpinState2020, oakesFastHighfidelitySingleshot2022}, we obtain an average generation rate of $g_{Bell} \approx 0.14$ MHz.

As we discussed in the main text, for smaller values of $T_{2, c}$, one needs to perform four rounds of purification. On average, this corresponds to $N_{avg} \approx 25.4$ and $33.0$ raw Bell pairs to successfully go through the four rounds, leading to average generation rates of $g_{Bell} \approx 2.6$ MHz and $g_{Bell} \approx 2.0$ MHz ($g_{Bell} \approx 24.9$ KHz and $g_{Bell} \approx 20.8$ KHz by incorporating gate time and measurement time) for $T_{2,c} = 100 $~\SI{}{\nano\second} and $T_{2,c} = 50 $~\SI{}{\nano\second} respectively.

\section{Shuttling}\label{sec:shuttling}

\subsection{Model Parameters and Details}

For all shuttling simulations, a Zeeman splitting of \SI{40}{\micro\eV} was used, corresponding to an external magnetic field strength of approximately \SI{350}{\milli\tesla} which is comparable to many silicon spin experiments. The inhomogeneous magnetic field \textbf{b} may be decomposed as a tranvserse gradient $b_x$ and a longitudinal gradient $b_z$ with no loss of generality.
In order to minimise coherent spin flips, shuttling should take place along an axis with minimal $b_x$.
As field gradients on the micron scale should be nearly constant, we assume the total inhomogeneity is evenly distributed down the shuttling channel with electrons only moving in a linear path and we consider $b_x=$~\SI{1}{\micro\eV}, $b_z=$~\SI{3}{\micro\eV}.

The detuning sweep over $t\in[0,2\epsilon_0/\alpha]$ is taken to be $\epsilon(t)=\alpha t - \epsilon_0$. $\epsilon_0=$~\SI{800}{\micro\eV} and $\alpha=$~\SI{300}{\micro\eV/\nano\second} are used to be similar with previous experiment \cite{millsShuttlingSingleCharge2019}. Intrinsic spin orbit coupling is included with a vector $\mathbf{\Omega} = (E_{\mathrm{SOI}},-E_{\mathrm{SOI}},0)$ for shuttling along the $[110]$ crystallographic axis of silicon \cite{ginzelSpinShuttlingSilicon2020}. The energy scale $E_{\mathrm{SOI}}$ is of the order \SI{0.1}{\micro\eV} as estimated from experiment \cite{yangSpinvalleyLifetimesSilicon2013}. We consider $E_{\mathrm{SOI}}=$~\SI{1}{\micro\eV} to be conservative.

While large valley splitting variation may be expected in devices at present, substantial effort is being placed on improving heterostructure quality such that future devices may have consistently large ground state gaps. In order to emphasize the importance of the valley phase parameter, we assume consistently high valley splittings by randomly generating them from a normal distribution with mean $|\Delta_D|=$~\SI{75}{\micro\eV} and standard deviation $SD_{|\Delta|}=$~\SI{10}{\micro\eV}. Such a case should apply to both Si/SiGe and Si-MOS devices. We focus on tunnel couplings $t_c$ larger than \SI{20}{\micro\eV}, as these have been experimentally reported for silicon charge shuttling \cite{millsShuttlingSingleCharge2019}. A tunnel coupling of order \SI{100}{\micro\eV} was reported in \cite{yonedaCoherentSpinQubit2021}, suggesting that even larger values are realistic.

In order to simulate the evolving entanglement between two spins labelled using $1$ and $2$, a second spin is added to the Hamiltonian of \cref{eqn:H_Si_DQD_global}:

\begin{equation}
    \label{eq:H_SSVO}
    \tilde{H} = H'_1 + \frac{B_2}{2} \otimes s_{z,2}.
\end{equation}

\noindent The Zeeman splitting of the stationary spin, $B_2$, is inconsequential, as we may work in the rotating frame of the second spin such that its dynamics are trivial. We assume that there is negligible residual exchange interaction between the two spins, and Coulomb interaction between electron charges may be accounted for separately.

The shuttled electron always begins in the ground valley-orbit state $\ket{L-}$. After the detuning sweep has completed, a following fast, deep detuning pulse is implicitly assumed, such that deterministic charge transfer is assured. As described in the main text, any population in the excited $\ket{L}$ states will evolve into the target dot. We assume that this secondary charge transfer is adiabatic, and that subsequent relaxation from excited orbital states preserves both spin and valley, such that the operation is described by the partial trace of the orbital degree of freedom. We point out that different assumptions could be taken at the expense of introducing additional microscopic parameters into the model. The spin-valley state $\ket{sv}$ is then re-initialised as $\ket{Lsv}$ prior to the next detuning sweep.

\subsection{Valley Projective Measurement}

For our analysis, we assume the spin-valley state is well defined throughout the shuttling protocol. In order to fit the protocol within a larger spin-based quantum computation, we propose projecting the valley state into the ground valley state. We make use of an ancilla dot which is populated with a single electron initialized in the lowest-energy state $\ket{-\downarrow}$, and describe the total two-electron two-dot system with the Hamiltonian:

\noindent Once again, $\epsilon$ describes the detuning between dots, $U$ is the Coulomb repulsion energy attributed to adding a second electron to a single dot, $t$ is a spin- and valley-preserving tunnel coupling between dots, $E_{Z,D}$ and $E_{V,D}$ are Zeeman and Valley splittings. $\hat{n}_i = \hat{c}_i^\dagger\hat{c}_i$ is the number operator for state $i$, while $\hat{c}_i^\dagger$ and $\hat{c}_i$ are creation and annihilation operators, respectively. Excluded indices are assumed to be summed over.

\cref{fig:valley_measure} illustrates how the ancilla dot may be used to implement a valley-to-charge conversion, such that the spin state of the ground valley is unaffected after post-selection.  We neglect the valley phase here as analyzing its effect in readout is not the present focus, although it certainly will affect readout quality \cite{tagliaferriImpactValleyPhase2018}. We make use of \cref{eq:H_2eDQD} to illustrate the essential physics and plot the corresponding energies in \cref{fig:energy_levels}. We note that this is only one possible choice. Information may be better preserved by coherently manipulating the valley state directly, or it may be further damaged through the relaxation of the valley state during the shuttling operation.

\begin{widetext}

\begin{align}
    \label{eq:H_2eDQD}
    H_{2e}  = \frac{\epsilon}{2}(\hat{n}_L - \hat{n}_R) &+ \frac{U}{2}\sum_{D\in\{L,R\}}\hat{n}_D(\hat{n}_D-1) 
     + t\sum_{v,s}(\hat{c}_{Rvs}^\dagger\hat{c}_{Lvs} + \hat{c}_{Lvs}^\dagger\hat{c}_{Rvs}) \\
    & + \sum_{D\in\{L,R\}}\frac{E_{Z,D}}{2}(\hat{n}_{D,\uparrow}-\hat{n}_{D,\downarrow}) + 
    \sum_{D\in\{L,R\}}\frac{E_{V,D}}{2}(\hat{n}_{D,+}-\hat{n}_{D,-}).
    \nonumber
\end{align}
\end{widetext}

\begin{figure*}[th]
	\begin{centering}
		\includegraphics[width=\textwidth]{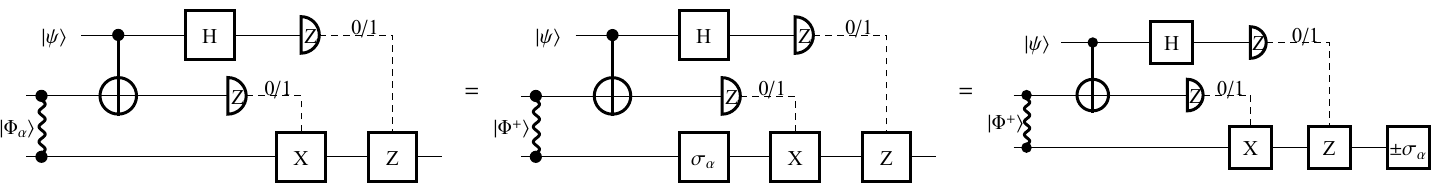}
		\caption{
			Quantum teleportation with injecting an other Bell pair than the required $|\Phi^+\rangle$ is equivalent to applying a Pauli operator $\sigma_\alpha$ to the teleported qubit state as $\pm \sigma_\alpha |\psi\rangle$ where the Pauli operator maps between the two Bell states as $|\Phi_\alpha\rangle := \sigma_\alpha |\Phi^+\rangle$. Due to linearity, using an imperfect, incoherent mixture of input Bell states for teleportation is equivalent to applying a Pauli noise channel to the teleported single-qubit state. Additionally applying twirling guarantees that this error channel is a single-qubit depolarising channel.
			\label{fig:teleportation_error}
		}
	\end{centering}
\end{figure*}

\subsection{Other Sources of Decoherence}

For this analysis into electron shuttling, we have focused on the intrinsic ability of the unitary tunnelling operation to populate excited valley-orbit states and thereby give rise to information loss. However, other external sources may also damage information. This includes the hyperfine coupling of the electron spins to residual $^{29}$Si nuclear spins, residual coupling to reservoirs, and ubiquitous $1/f$ charge noise. We expect the effect of hyperfine dephasing to be minimal, since the timescale of shuttling is orders of magnitude smaller than 10-100~\SI{}{\micro\second} $T_2^*$ times reported in purified silicon \cite{veldhorstTwoqubitLogicGate2015}. Similarly, coupling to reservoirs may require hundreds of tunnelling events to have a measurable effect \cite{ginzelSpinShuttlingSilicon2020}.

To quantify the effect of charge noise, we adapt the technique used in \cite{krzywdaAdiabaticElectronCharge2020} by adding $1/f$ charge noise to the detuning sweep $\epsilon(t)$ generated by combining 1000 Ornstein-Uhlenbeck processes and averaging over 100 cases. The power spectrum is normalized with respect to the $S$(\SI{1}{\mega\hertz})$=10^{-6}$~\SI{}{\micro\eV^2/\hertz} reported in the charge noise spectroscopy of a present-day Si/SiGe heterostructure \cite{connorsChargenoiseSpectroscopySi2022}. From \cref{fig:shuttling_charge_noise}, we can see that this magnitude of charge noise adds a negligible correction to the anticipated unpurified and purified concurrence estimates. We note that charge noise may play a much larger role when using detuning sweep speeds substantially slower than \SI{300}{\micro\eV/\nano\second} or tunnel couplings smaller than \SI{10}{\micro\eV}.

\section{Error Suppression}
\subsection{Quantum teleportation with imperfect Bell pairs \label{sec:proof_pauli_error}}

Let us first show the following useful identity. Quantum state teleportation can be realised by consuming a Bell pair $\ket{\Phi^+} := (|00\rangle + |11\rangle)/\sqrt{2}$. If we instead inject any one of the three other Bell pairs it will result in a Pauli operation on the teleported qubit. In particular, we can define the four Bell states in terms of Pauli transformations as $|\Phi_\alpha\rangle := \sigma_\alpha |\Phi^+\rangle$ where $\sigma_\alpha$ are Pauli matrices with $\alpha \in \{0,1,2,3\}$. It is now straightforward to show the identity in \cref{fig:teleportation_error}

It immediately follows from linearity that injecting an incoherent mixture of Bell pairs with inhomogeneous weights $p_\alpha$ (probabilities) as
\begin{equation} \label{eq:bell_error}
    \Phi_{mixture} := \sum_{\alpha=0}^3 p_\alpha |\Phi_\alpha \rangle  \langle \Phi_\alpha|
\end{equation}
and performing the teleportation results in a qubit state that has effectively undergone an inhomogeneous Pauli error channel as
\begin{equation*}
    \mathrm{Pauli}(\rho) :=
    (1- p_0)\rho
    +p_1 X  \rho X
    +p_2 Y \rho Y
    +p_3 Z \rho Z.
\end{equation*}

It also follows from linearity that if we inject a coherent superposition of Bell states
\begin{equation*}
    |\Phi_{coherent}\rangle
    :=
    \sum_{\alpha=0}^3 c_\alpha
    |\Phi_\alpha \rangle 
\end{equation*}
then it shows up as a coherent transformation of the final state as $U |\psi\rangle$.

One might additionally apply twirling techniques to the Bell pairs such that the output state is a Werner state \cite{wernerQuantumStatesEinsteinPodolskyRosen1989}, i.e., all three erroneous Bell pairs appear with identical probabilities resulting in an effective single-qubit depolarising channel acting on the teleported qubit state. These twirling techniques randomly apply Pauli operators $\sigma_\alpha$ to the input Bell pair such that the ideal state $|\Phi^+\rangle$ is left invariant while the erroneous Bell states are mapped onto each other.

Since the ESD/VD error suppression techniques are oblivious to coherent errors, it is important that the prepared/distilled Bell states are of the form of \cref{eq:bell_error}. For this reason we require that twirling techniques \cite{wernerQuantumStatesEinsteinPodolskyRosen1989} are implemented:
This requires a minor overhead of applying local single-qubit operations randomly to the input Bell pairs but in return they guarantee that imperfections are incoherent as in \cref{eq:bell_error}. Consequently we can model the teleportation process implicitly as single qubit Pauli error channels applied to the teleported single-qubit states.

\subsection{Numerical simulations \label{app:num_sim}}

In \cref{fig:teleportation}(c) we simulate the spin-ring Hamiltonian in \cref{spin_ring} and aim to determine its ground state
using the Variational Hamiltonian Ansatz: this ansatz consist of alternating layers of time evolutions under the Hamiltonians which we define as
\begin{equation*}
	\mathcal{H}_{0} :=  \sum_{k = 1 }^N \omega_k Z_k
	\quad \quad \text{and}  \quad \quad 
	\mathcal{H}_{1} := J \, \sum_{k \in \text{ring}(N)}  \vec{\sigma}_k \cdot \vec{\sigma}_{k+1},
\end{equation*}
via $\mathcal{H} = \mathcal{H}_0 + \mathcal{H}_1$. We start the optimization from the ground state $|\psi_{init}\rangle$ of the diagonal Hamiltonian $\mathcal{H}_{0}$ which we have analytically determined \cite{koczorExponentialErrorSuppression2021}. We then apply alternating layers
of the parametrised evolutions $A(\gamma_k):= e^{-i \gamma_k \mathcal{H}_1 }$ and $B(\beta_k):=  e^{-i \beta_k \mathcal{H}_0 }$
to this initial state as
\begin{equation*}
	|\psi(\underline{\beta}, \underline{\gamma}) \rangle =  
	 B(\beta_l)  A(\gamma_l)  \dots A(\gamma_2) B(\beta_1)  A(\gamma_1)  \,	|\psi_{init}\rangle,
\end{equation*}
using overall $l$ layers. In typical applications the parameters $\underline{\beta}$ and $\underline{\gamma}$ are optimised by a
classical co-processor such that the experimentally estimated energy $E:=\tr[\rho \mathcal{H}]$ is minimised. 

We simulate a quantum device that can natively implement controlled-$Z$ gates between any pairs of qubits as well as single qubits. As such, we remark that we do not take into account the additional complexity imposed by connectivity constraints: one needs to apply SWAP operations to overcome limited connectivity. However, these operations are part of the main computation and their imperfections are fully mitigated by the derangement circuits. Furthermore, we expect their complexity can be negligible when compared to the complexity of the main computation (almost all terms in our Hamiltonian are nearest-neighbour interactions).

We have determined the ground state energy to a precision $\Delta E = 10^{-4}$ using $l=20$ layers of the ansatz by optimising parameters in a noise-free scenrio. We use these parameters as input for our noisy simulations. In particular, we assume a noise model where single-qubit gates are followed by single-qubit depolarising events, while 2-qubit gates are followed by 2-qubit depolarisation. The gate error of 2-qubit gates is taken $5$-times larger as of single-qubit gates as explained in the main text.


%

\end{document}